# Self-potentials in partially saturated media: the importance of explicit modeling of electrode effects

D. Jougnot[1] and N. Linde[1]

[1]Applied and Environmental Geophysics Group, University of Lausanne, Lausanne, Switzerland.

**Authors contact:**
Damien Jougnot:
damien.jougnot@unil.ch

Niklas Linde:
niklas.linde@unil.ch

**Abstract:**

Self-potential (SP) data are of interest to vadose zone hydrology because of their direct sensitivity to water flow and ionic transport. There is unfortunately little consensus in the literature about how to best model SP data under partially saturated conditions and different approaches (often supported by one laboratory data set alone) have been proposed. We argue herein that this lack of agreement can largely be traced to electrode effects that have not been properly taken into account. A series of drainage and imbibition experiments are considered, in which we find that previously proposed approaches to remove electrode effects are unlikely to provide adequate corrections. Instead, we explicitly model the electrode effects together with classical SP contributions using a flow and transport model. The simulated data agree overall with the observed SP signals and allow decomposing the different signal contributions to analyze them separately. By reviewing other published experimental data, we suggest that most of them include electrode effects that have not been properly taken into account. Our results suggest that previously presented SP theory works well when considering the modeling uncertainties presently associated with electrode effects. Additional work is warranted to not only develop suitable electrodes for laboratory experiments, but also to assure that associated electrode effects that appear inevitable in longer-term experiments are predictable, such that they can be incorporated in the modeling framework.

# 1. Introduction

The self-potential (SP) method is of direct interest to vadose zone hydrology as SP data are sensitive to water fluxes under partially saturated conditions. The SP method is passive and consists in measuring naturally occurring electrical potential differences using non-polarizable electrodes in or at the surface of geologic media. This electrical potential distribution is governed by a Poisson equation with source terms related to the divergence of source current densities (Sill, 1983). The sources have three main origins that are related to water flow, ionic diffusion, and redox phenomena. The different source types have been studied at different scales and with different degrees of control, including both laboratory and field studies. The underlying physical processes are overall well understood, but predictive modeling is challenging given the typically incomplete knowledge about the porous media and the pore water under question.

The source currents density related to water flow are of electrokinetic (EK) nature and are called streaming currents, while the resulting component of the SP signal is the streaming potential. The electrical charge at the mineral surface of geological porous media is associated with an electrical double layer. When water flows through such a medium, the excess charge in the diffuse layer is dragged, creating thereby a streaming current. Early studies under two phase flow conditions indicated that the SP method could be used to sense water fluxes in both laboratory (Antraygues and Aubert, 1993; Sprunt et al., 1994) and field conditions (Thony et al. 1997). Later, many studies were carried out to improve models that predict the relative voltage coupling coefficient, which relates electrical potential and hydraulic pressure differences at a given saturation divided by the value at full saturation. Its behavior as a function of saturation is non-linear and medium dependent (e.g., Guichet et al., 2003; Revil and Cerepi, 2004; Allègre et al., 2010; Vinogradov & Jackson, 2011). Instead of studying the relative voltage coupling coefficient, other studies focus directly on how the corresponding effective excess charge varies as a function of saturation (Linde et al., 2007; Revil et al., 2007; Mboh et al., 2012; Jougnot et al., 2012). This is advantageous as the relative voltage coupling coefficient is an aggregate parameter, which not only depends on the relative effective excess charge, but also on the relative permeability and electrical conductivity functions (Linde et al., 2007). Previous laboratory investigations have mainly focused on drainage experiments (Linde et al., 2007; Allègre et al., 2010; Mboh et al., 2012). Only two studies present imbibition experiments, but without any associated modeling: Haas and Revil (2009) focus on the electrical burst signature, while Vinogradov and Jackson (2011) measured the voltage coupling coefficient at different saturations in gas/brine and oil/brine systems. All these works are motivated by the possibility of using SP data to quantitatively investigate water fluxes in different types of partially saturated media: unconsolidated sediments, natural soils, reservoir rocks, and even in melting snow (Kulessa et al., 2012). A non-invasive and reliable data source that is directly related to filtration and evaporative fluxes in soils would clearly be of great use in vadose zone hydrology. Even if SP source generation in geological media is fairly well understood, more detailed experimental studies are needed to test or develop interpretative frameworks prior to quantitative field applications (see Linde et al., 2011). This could refer to well-controlled laboratory studies or intermediate scale investigations (e.g., lysimeter studies; Doussan et al., 2002).

One important and often over-looked aspect in previous studies is that other possible contributions to the measured SP data have to be considered and possibly removed before interpreting SP measurements in terms of water flow. In fact, other contributions sum up with the streaming potential and may be as large or even larger in magnitude. The contribution of ionic diffusion to the SP signal is linked to the charge separation along the activity gradients due to different ionic mobilities in the pore water. This phenomenon has been studied experimentally in saturated conditions (e.g., Maineult et al., 2004, 2005; Revil et al. 2005), but not in partially saturated media. A theoretical model for two-phase conditions can be found in Revil and Jougnot (2008). Woodruff et al (2010) explain SP data in boreholes by considering the ionic diffusion contribution in a brine/oil system. Other contributions to the SP signal include sources related to redox processes. They have recently become the subject of an increasing number of studies focusing on delimitating and characterizing contaminated sites (e.g., Naudet et al., 2003; Maineult et al., 2006; Linde and Revil, 2007; Zhang et al., 2010). Contributions related to redox processes can only occur when an electronic conductor is available that connects the reductive and oxidative part of the medium (e.g., a metallic pipe as in Castermant et al. (2008) or possibly connected biofilms), and appears rather uncommon even in controlled laboratory conditions (see Hubbard et al., 2011) or in contaminated sites (e.g., Slater et al., 2010). This effect is not important for the example considered herein and we refer the reader to these references for further information.

The equilibrium electrode potential (i.e., the electrical potential of an electrode measured against a reference electrode when there is no current flowing through the electrode) is another concern and somewhat ignored problem in the SP literature. This effect has in our mind hindered the development of SP as a quantitative method for vadose zone investigations and created significant discrepancies between the experimental data obtained by different research groups (e.g., Linde et al. (2007) and Mboh et al. (2012) on the one hand, and Allègre et al. (2010) on the other hand). Self-potential measurements are usually conducted with so-called non-polarizable electrodes, that is, electrodes whose electrode potential is not significantly affected by the current passing through them as the electrode reaction is inherently fast. These electrodes can be made of various metal-salt couples, where the salts are either solid or in solution. But, even if the current is not affecting the electrode potential significantly, other phenomena may influence the equilibrium electrode potential: namely chemistry and temperature.

The equilibrium electrode potential problem has been studied for field-based measurements and electrodes, but appears to have been largely ignored in recent SP laboratory studies. Petiau and Dupis (1980) studied noise and stability of various electrodes: among others, Pb-$PbCl_2$, Cu-$CuSO_4$, and Ag-AgCl. This work resulted in the design of the very stable Pb-$PbCl_2$ field electrode by Petiau (2000), in which the electrode is inserted in a chamber filled with brine-saturated clay that keeps the chemical conditions in the vicinity of the electrode very stable over time (see also Junge et al. (1990) for Ag-AgCl electrodes). For long-term laboratory measurements, a major problem arises with this design, namely the leakage of the chamber fluid into the investigated media, which could impact the medium and strongly affect the results. To minimize this problem, Maineult et al. (2004) developed laboratory-scale Cu-$CuSO_4$ electrodes, in which electrode leakage was decreased by choosing relatively thick walls of the porous pot.

Ag-AgCl electrodes are widely used in biology and electrochemistry, where they are also used as reference electrodes (Janz and Ives, 1968). Based on their study, Petiau and Dupis (1980) found Ag-AgCl electrodes to be among the best non-polarizable electrodes for geophysical applications, given their low noise and short stabilization time, but other works mention important electrode effects. For example, Antraygues and Aubert (1993) consider that these electrodes cannot be used for SP studies because of strong potential drift and high noise level, although no details can be found about the type of Ag-AgCl electrodes used in this study (sintered or not). Zhang et al. (2010) used a dual electrode design to remove SP contributions in a study of electrodic potential (EP) during microbial sulfate reduction (Ag/AgCl metal as sensing/EP electrode, Ag/AgCl metal in KCl gel as reference/SP electrode).

In partially saturated media, Linde et al. (2007), Allègre et al. (2010), and Mboh et al. (2012) present significant residual SP values at the end of their experiments that appear unrelated to water flux. Self-potential data are usually interpreted after some sort of simplistic removal of electrode effects such as removing linear drifts of the electrodes between two points in time when the true SP signals are assumed to be known (e.g., Suski et al., 2004; Mboh et al., 2012) or by shifting the raw data to zero after a monitored hydrologic event (e.g. Linde et al., 2007). One source of electrode potential differences are temperature differences between the sensing and the reference electrodes. This problem has been studied for field electrodes and the choice of an appropriate pH of the brine allowed Petiau (2000) to minimize this effect (see also Hubbard et al., 2011). Indeed, even diurnal variations in room temperature can lead to significant effects on SP data (e.g., Maineult et al., 2004, 2005; Allègre et al., 2010). These effects are often neglected, as it is difficult to remove these contributions from the data.

In this work, we monitor SP signals during repeated drainage and imbibition cycles in a sand column. We then model the different contributions to the measured SP data including those arising from the porous media itself and those related to the electrodes. We demonstrate that diffusion-related current sources and electrode effects must be included to enable quantitative analysis of SP monitoring experiments under partially saturated conditions. Previously proposed electrode corrections are found to be insufficient for detailed analysis of even short-term experiments lasting a few hours (Linde et al., 2007; Mboh et al., 2012). Even if the experimental details vary, we argue that electrode effects are present in most published experimental SP data acquired during drainage experiments. To advance the petrophysical basis of the SP method and to make it a reliable tool to infer in situ water flux, we argue based on the results presented herein that careful consideration of electrode effects and their subsequent removal are of primary importance. If not, there is a clear risk that each new experimental data set that is affected by unaccounted or incorrectly filtered electrode effects is used to propose new and possibly biased theory.

## 2. Theory

### *2.1 Electrode design and equilibrium potentials*

The SP method consists of measuring naturally occurring electrical potential differences in a geological medium with respect to a given reference electrode. Self-potential measurements can only be made when two electrodes (potential and reference) are linked by an electrically conductive medium. To assure high-quality data, the voltmeter used should have an internal resistance that is many orders of magnitudes larger than the electrical resistance of the Earth circuit connecting the electrodes. In this section, we describe the design and operation of the electrode-geological system, which is also referred to as an electrochemical cell.

Electrochemical cells can either be used to generate a current arising from chemical reactions, or to induce a chemical reaction by current flow. Each half-cell of the system consists of an electrode and an electrolyte (either the solution found in a chamber around the electrode or the pore water of the geological medium). Electrodes are electrical conductors that ensure the electrical contact with the nonmetallic part of the circuit. An electrode is called a cathode if electrons leave the cell (oxidation) and anode if electrons enter the cell (reduction). Electrodes are classified into those of the first and second kind (Janata, 2009).

Electrodes of the first kind can be either cationic or anionic. In both cases, equilibrium is obtained between a chemical element (or a molecule) and the corresponding ions in solution. For SP measurements, only cationic electrodes are used and, for a given metal M, equilibrium is established with the corresponding cation $M^{+n}$,

$$M(s) \rightleftharpoons M^{+n} + ne^-. \qquad [1]$$

In the above chemical reaction, $n$ is the valence of the cation, $e^-$ represents an electron, and (s) states that the element M is in its solid phase. The electrode potential is given by the Nernst equation,

$$\varphi^{elec} = \varphi^0_M + \frac{k_B T}{n e_0} \ln\{M^{+n}\}, \qquad [2]$$

where $\varphi^{elec}$ and $\varphi^0_M$ are the electrode and the standard electrode potentials (V), respectively, $e_0 = 1.6 \times 10^{-19}$ C is the elementary charge, $k_B = 1.3806 \times 10^{-23}$ J K$^{-1}$ the Boltzmann constant, $T$ the absolute temperature (K) and $\{M^{+n}\}$ the cation activity (-). It is thus clear that electrical potentials of electrodes of the first kind are directly related to the temperature and the cation activity of the solution.

Figure 1 displays a schematic view of an electrode of the second kind. The solid part consists of a metal M(s) covered by a layer of its soluble salt $MX_n(s)$ that is immersed in a solution of a soluble salt containing the anion $X^-$. The soluble salt covering the metal can be dissolved partially in the solution following:

$$M^{+n} + nX^- \rightleftharpoons MX_n(s). \qquad [3]$$

This dissolution is controlled by $K_{MX_n} = \{M^{+n}\}\{X^-\}^n$, the solubility product of the salt $MX_n(s)$. It yields an equilibrium between the metal atoms and the solution anions through the two partial equilibria described in Eq. [1] and [3]:

$$M(s) + nX^- \rightleftharpoons MX_n(s) + ne^-. \qquad [4]$$

The solution can be split into two parts: the direct vicinity of the electrode-solution interface and the bulk solution. We define the bulk solution as all electrolytes in between two electrodes (e.g., the pore water of the geologic medium in which the SP measurements are carried out). Ionic transfers between the two regions occur naturally by diffusion, convection, and ionic migration (Agar, 1947).

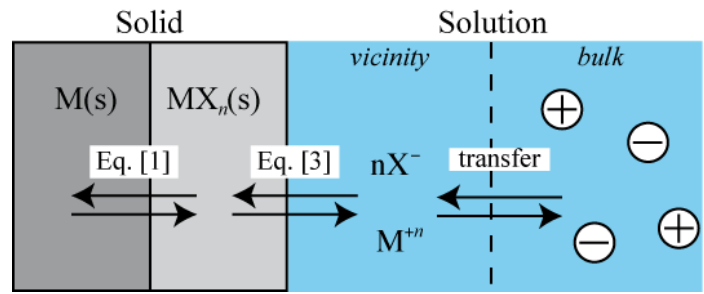

*Fig. 1. Sketch of an electrode of the second kind M-$MX_n$.*

If $K_{MX_n}$ is sufficiently low, it maintains a constant activity of the cation $\{M^{+n}\}$ at the electrodes. From Eq. [1] and [3], the electrical potential of an electrode of the second kind can be expressed by the modified Nernst equation (Janata, 2009):

$$\varphi^{elec} = \varphi_M^0 + \frac{k_B T}{ne_0} \ln K_{MX_n} - \frac{k_B T}{e_0} \ln\{X^-\}. \qquad [5]$$

It is seen that the temperature and the anion activity {X⁻} in the surrounding solution play important roles in determining the electrode potential $\varphi^{elec}$ (Janz and Taniguchi, 1953; Raynauld and Laviolette, 1987).

Self-potential measurements are mainly carried out using three different types of electrodes: Pb-PbCl$_2$ (e.g., Petiau and Dupis, 1980; Petiau, 2000, Kulessa et al., 2012), Cu-CuSO$_4$ (e.g., Maineult et al., 2004; 2005; 2006), and Ag-AgCl (e.g., Junge, 1990; Linde et al., 2007; Allègre et al., 2010; Zhang et al., 2010; Mboh et al., 2012). In many of these studies, the electrodes are placed in a chamber filled with a solution that is separated from the geologic medium by a porous medium (ceramic or wood).

We consider hereafter Ag-AgCl electrodes that are used in the experimental study presented below. We start by quantifying the effect of temperature and chemical activities upon their equilibrium electrode potential and the resulting effect on the SP measurements. The reaction at the electrode is,

$$\text{Ag(s)} + \text{Cl}^- \rightleftharpoons \text{AgCl(s)} + e^-. \qquad [6]$$

From this reaction it is clear that the solution in contact with the Ag-AgCl electrode has to contain chloride ions to function properly. If this is not the case, the electrodes will dissolve and create irreversible reactions (Raynauld and Laviolette, 1986). Janz and Ives (1968) write: "It is equally important that the solution phase shall contain an adequate concentration of chloride ion-otherwise, the electrode potential becomes unstable, irreproducible, and not theoretically significant, lying at the mercy of dissolved oxygen, or other adventitious impurities". These concerns are very strong arguments against using chamber electrodes filled with distilled water (e.g., Guichet et al., 2003; Allègre et al., 2010). The very low solubility product of AgCl ($K_{AgCl} = 1.8 \times 10^{-10}$ at 25°C) fulfills the necessary condition to use Eq. [5] (i.e., the Ag+ activity at the electrode is constant: Janz and Ives, 1968; Janata, 2009). The standard potential of the Ag-AgCl electrode ($\varphi^0_{Ag\text{-}AgCl}$) depends on the standard potential of silver ($\varphi^0_{Ag}$ = 0.799 V) and the temperature (Fig. 2). It is clear that the temperature has a fairly important influence on the electrode standard potential (~ 40 mV for a 60 °C change). It is then possible to directly link the equilibrium electrode potential with the temperature and chloride activity (Gilbert, 1947):

$$\varphi^{elec} = \varphi^0_{Ag\text{-}AgCl}(T) - \frac{k_B T}{e_0} \ln\{Cl^-\} \,. \qquad [7]$$

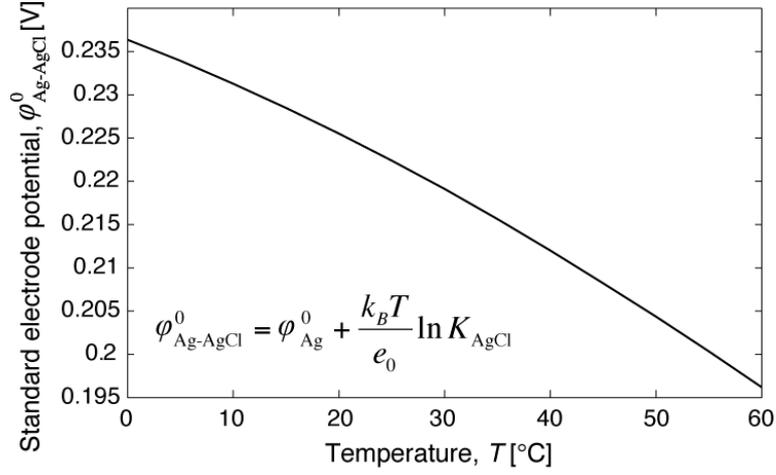

*Fig. 2. The standard Ag-AgCl electrode potential $\varphi^0_{Ag\text{-}AgCl}$ with respect to temperature.*

To understand the importance of Eq. [7] on actual SP measurements, one has to consider both the potential and reference electrodes, hereafter referred to with indices *i* and *ref*, respectively. We distinguish between SP signals associated with the geological medium $\varphi_i$ and those due to the electrode potential $\varphi_i^{elec}$ (electrode effects). This distinction can be expressed as (Linde et al., 2011):

$$SP_i^{meas} = (\varphi_i + \varphi_i^{elec}) - (\varphi_{ref} + \varphi_{ref}^{elec}) \,. \qquad [8]$$

From Eq. [7] and [8], the equilibrium electrode potential difference contribution, $SP_i^{elec} = \varphi_i^{elec} - \varphi_{ref}^{elec}$, to the measured SP signal can be expressed as:

$$SP_i^{elec} = \left(\varphi^0_{Ag\text{-}AgCl}(T^i) - \frac{k_B T^i}{e_0}\ln\{Cl^-\}_i\right) - \left(\varphi^0_{Ag\text{-}AgCl}(T^{ref}) - \frac{k_B T^{ref}}{e_0}\ln\{Cl^-\}_{ref}\right). \qquad [9]$$

Equation [9] illustrates that it is the differences in temperature and/or chloride activity in the vicinity of the $i^{th}$ electrode with respect to the reference electrode that matters. Electrode effects calculated from this equation as a function of temperature and chloride activity are of significant amplitude. For example, if the temperature and chemical activity in the vicinity of the reference electrode are $T^{ref} = 25°C$ and $\{Cl^-\}_{ref} = 10^{-4}$, while they are $T^i = 22°C$ and $\{Cl^-\}_i = 0.5 \times 10^{-4}$ in the vicinity of the $i^{th}$ potential electrode, the resulting electrode potential is $SP_i^{elec} = 17.12$ mV. A temperature difference of 3°C could easily be reached in laboratory conditions (e.g., temperatures inside and outside the experimental column of Allègre et al. (2010), see their appendix A), while an activity ratio of 0.5 (or more) could occur especially at low chemical activity due to local concentration gradients caused by chemical reactions (mineral dissolution) or by the change of the pore water (infiltration). If the temperatures at the two electrodes are the same ($T^i = T^{ref}$), Eq. [9] can be simplified to

$$SP_i^{elec} = -\frac{k_B T^i}{e_0} \ln \frac{\{Cl^-\}_i}{\{Cl^-\}_{ref}}. \qquad [10]$$

This high sensitivity to chloride activity has led some authors to use Ag-AgCl electrodes to measure chloride activity directly, for example, in cement based materials (Elsener et al., 2003).

We have highlighted above that chloride ions in solution are necessary for Ag-AgCl electrodes to function properly and that their activities in the vicinity of the electrodes have to be the same to avoid electrode effects (i.e., equilibrium electrode potential differences). If this is not the case, it is important that $\{Cl^-\}$ stays either constant or that its evolution over time can be predicted or monitored. Electrodes within chambers that are separated from the geological medium by a porous ceramic or wood allow electrical contact, but limits ionic exchange. This helps to stabilize the anion activity within the chamber and hence in the vicinity of the electrode-solution interface. In the field of neurosciences or physiology, where measured electrical potential amplitudes may be a few tenths of µV (typically between 10 and 100 µV in electroencephalography on the scalp; e.g., Aurlien et al., 2004), the problems of chemical activity-related signals have been the subject of several studies and proposed electrode designs (e.g., Snyder et al., 1999; Shao and Feldman, 2007). For these types of experiments, Ag-AgCl electrodes are usually plunged in a saline gel with high chloride ion concentrations that buffer against activity variations (e.g. Tallgreen et al., 2005).

The use of chamber electrodes can be problematic in SP studies in geological media since they induce two types of perturbations: electrode leakage into the geologic medium and diffusion potential. The diffusion potential around electrodes (often called junction potential) arises due to the differential diffusion of ions with different mobilities across the porous chamber material. The diffusion phenomena could either be from the electrodes (e.g., chamber filled with a high concentration of Cl⁻ gel) or toward the electrodes (e.g., chamber filled with deionized water). This phenomenon generates an electrical signal that is an additional perturbation to the measurements (Barry and Diamond, 1970). To avoid these phenomena in laboratory SP studies, some authors use the same water in the chambers as in the pore water (e.g., Mboh et al., 2012).

*2.2 Self-Potential modeling framework*

A soil with a porosity $\phi$ can be occupied by different volume fractions of air and water. The water saturation $S_w$ of the medium corresponds to the ratio between the water and pore volumes. The effective water saturation is defined by,

$$S_e = \frac{\theta_w - \theta_w^r}{\phi - \theta_w^r}, \qquad [11]$$

where $\theta_w$ is the volumetric water content defined by: $\theta_w = S_w \phi$, and $\theta_w^r$ is the residual water content. The saturation of the medium depends on the water pressure $p_w$ (Pa) and we use the van Genuchten model (van Genuchten, 1980) to describe this dependence:

$$S_e = \left[1 + \left(\alpha_{VG} p_w\right)^{n_{VG}}\right]^{-m_{VG}}, \qquad [12]$$

where $\alpha_{VG}$ (Pa$^{-1}$) corresponds to the inverse of the air-entry pressure, while $n_{VG}$ and $m_{VG} = 1 - (1/n_{VG})$ are curve shape parameters.

The vertical water flux $u$ (m s$^{-1}$) distribution in the column during the different hydrological events is modeled through the 1D Richards' equation using

$$u = -\frac{K_w(S_e)}{\rho_w g} \frac{\partial}{\partial z}\left(p_w - \rho_w g z\right), \qquad [13]$$

where $\rho_w$ is water density (kg m$^{-3}$), $g$ the gravitational acceleration (9.81 m s$^{-2}$), z the elevation (m) and $K_w(S_e)$ the hydraulic conductivity. This last parameter is a function of saturation and can be described by the van Genuchten-Mualem model:

$$K_w(S_e) = K_w^{sat} \sqrt{S_e} \left[1 - \left(1 - S_e^{1/m_{VG}}\right)^{m_{VG}}\right]^2, \qquad [14]$$

with $K_w^{sat}$ the hydraulic conductivity in saturated conditions, which is related to the intrinsic permeability, $k$ (m$^2$), of the medium by,

$$k = \frac{K_w^{sat} \eta_w}{\rho_w g}, \qquad [15]$$

where $\eta_w$ is the dynamic viscosity of the water ($\eta_w = 10^{-3}$ Pa s$^{-1}$ at 25°C).

Given the importance of the chloride activity upon the electrode effect, the ionic transport has to be simulated for a given hydraulic flow field. The water is an electrolyte containing $Q$ ionic species $j$ with a concentration $C_j$ in the water phase. Given the fairly low ionic concentrations used in this study, we consider that activities can be approximated by concentrations $C_j \approx \{j\}$ in the transport simulations. The transport under partial saturation is driven by the water flux through the 1D conservation equation:

$$\frac{\partial(\theta_w C_j)}{\partial t} + \frac{\partial}{\partial z}\left[-\left(\alpha_z \frac{u^2}{\theta_w |u|} + D_j^{eff}\right)\frac{\partial C_j}{\partial z} + u C_j\right] = 0, \qquad [16]$$

where $\alpha_z$ is the dispersivity of the medium along the $z$ axis, and $D_j^{eff}$ the effective diffusion coefficient of the medium (m$^2$ s$^{-1}$). To describe the evolution of $D_j^{eff}$ as a function of saturation, we use the model by Revil and Jougnot (2008). In this approach, petrophysical parameters obtained from electrical conductivity measurements of the medium are used to describe the geometry of the water phase within the pore space (the resulting information is similar to tortuosity and constrictivity; e.g., van Brakel and Heertjes, 1974). Hamamoto et al. (2010) applied this approach to various geological media under partially saturated conditions. Considering the relatively large pore width of sand (used in our experiments), it is possible to neglect the influence of the mineral surface charges upon the ionic diffusion and to use the following expression,

$$D_j^{eff} = \frac{S_w^n}{F} D_j^w, \qquad [17]$$

where $D_j^w$ is the ionic diffusion coefficient of $j$ in the pore water (m$^2$ s$^{-1}$), while $F$ and $n$ are the formation factor and Archie's saturation exponent, respectively. The value of $F$ can be obtained the porosity through $F = \phi^{-m}$, where $m$ is Archie's cementation exponent. From the chloride ion concentration distribution within the porous medium, the electrode potential differences $SP_i^{elec}$ of the $i^{th}$ electrode can be calculated using Eq. [9] by assuming $C_j \approx \{j\}$. We consider only isothermal conditions herein (Eq. [10]), which is possible if applying a temperature correction to the SP data.

The flow and transport simulation provide the necessary input parameters to solve the SP problem at all times. From Sill (1983), two equations describe the SP response of a given source current density $\mathbf{J}_s$ (A m$^{-2}$),

$$\mathbf{J} = \sigma \mathbf{E} + \mathbf{J}_S, \qquad [18]$$

$$\nabla \cdot \mathbf{J} = 0, \qquad [19]$$

where $\mathbf{J}$ is the total current density (A m$^{-2}$), $\sigma$ is the medium electrical conductivity (S m$^{-1}$), $\mathbf{E} = -\nabla\varphi$ is the electrical field (V m$^{-1}$), and $\varphi$ is the electrical potential (V). The source current densities can be understood as forcing terms that tend to perturb the system from electrical neutrality. This induces an electrical current that instantaneously re-establishes electrical neutrality and the SP response are the associated voltage differences related to this current. In the absence of external source currents, it is possible to combine Eq. [18] and [19] to obtain

$$\nabla \cdot (\sigma \nabla \varphi) = \nabla \cdot \mathbf{J}_S. \qquad [20]$$

The two main source types (electrokinetic, superscript EK, and electrodiffusive, superscript diff) can be summed to obtain the total source current density in the medium: $\mathbf{J}_S = \mathbf{J}_S^{EK} + \mathbf{J}_S^{diff}$.

The electrokinetic source ($\mathbf{J}_S^{EK}$) is directly related to the water flux. Many models attempt to explain this source term in partially saturated media. Herein, we use an approach based on the concept of excess charge density proposed by Revil and Leroy (2004) in water saturated conditions and later extended to partial saturation by Linde et al. (2007). The excess charge in the pore water is due to electrical charges at mineral surfaces and the creation of an electrical double layer at the mineral-solution interface. While the water flows in the pores, the excess charge is dragged in the medium and generates this current source density. A complete framework has been developed by Revil et al. (2007) and applied to other homogeneous media under partial saturation (e.g., Mboh et al. 2012). Jougnot et al. (2012) further developed this approach by using a conceptualization of the porous medium as a bundle of capillaries that enabled calculating the excess charge density effectively dragged in the pore medium as a function of saturation: $\overline{Q}_v^{eff}(S_w)$ (see also, Linde, 2009; Jackson, 2010; and Jackson and Leinov, 2012). Following Jougnot et al. (2012), a saturated capillary size distribution with respect to saturation is obtained either from the water retention (e.g., Eq. [12]) or the relative permeability function (e.g., Eq. [14]). The electrokinetic source current density is defined by:

$$\mathbf{J}_S^{EK} = \overline{Q}_v^{eff}(S_w)\mathbf{u}. \qquad [21]$$

The source current density induced by diffusive phenomena ($\mathbf{J}_S^{diff}$) is generated by the ionic charge separation between pore water anions and cations with different mobilities $\beta_j$. This effect is described through the microscopic Hittorf number of each ion $j$,

$$t_j^H = \frac{\beta_j}{\sum_{\iota=1}^{Q} \beta_\iota}. \quad [22]$$

This parameter represents the fraction of the total current transported by a given ionic species. The charge separation can either be amplified or decreased by the electrical double layer. For such cases, the porous medium acts like a membrane and one has to use a macroscopic Hittorf number $T_j^H$, which takes the electrical double layer into account (for partially saturated conditions, see Revil and Jougnot, 2008). The electrodiffusive source current density is given by,

$$\mathbf{J}_S^{diff} = -k_B T \sum_{j=1}^{Q} \left( \frac{T_j^H(S_w)}{q_j} \sigma(S_w) \right) \nabla \ln\{j\}, \quad [23]$$

where $q_j = \pm Z_j e_0$ is the electrical charge of the considered ions (C), with $Z_j$ its valence. The resulting electrical potential is referred to with different names: electrodiffusive and junction potential (e.g., Maineult et al., 2004; Jouniaux et al., 2009), or membrane potential if the effect of the electrical double layer cannot be neglected (e.g., Revil et al., 2005).

Given the relatively large pore width of the studied medium (i.e., sand) and the ionic concentrations used, the effect of the electrical double layer on the electrodiffusion can be safely neglected and Eq. [23] can be simplified using the microscopic Hittorf number $t_j^H$ instead of the macroscopic one ($T_j^H$). In 1D, the electrical problem can hence be simplified and re-written as:

$$\frac{\partial}{\partial z}\left(\sigma(S_w)\frac{\partial \varphi}{\partial z}\right) = \frac{\partial J_S}{\partial z}, \quad [24]$$

$$J_S = \overline{Q}_v^{eff}(S_w)u - k_B T \sum_{j=1}^{Q} \left( \frac{t_j^H}{q_j} \sigma(S_w) \right) \frac{1}{C_j}\frac{\partial C_j}{\partial z}. \quad [25]$$

From Eq. [24] and [25], it is clear that an accurate model describing the electrical conductivity distribution is needed. Comparing five pedo-electrical models, Laloy et al. (2011) found that the model of Linde et al. (2006), based on Pride's (1994) volume averaging approach, was the most suitable to describe the evolution of electrical conductivity in a partially saturated media (see Revil et al. (2007) and Revil and Jougnot (2008) for other applications). It is given by,

$$\sigma(S_w) = \frac{1}{F}\left[S_w^n \sigma_w + (F-1)\sigma_S\right], \quad [26]$$

where $\sigma_w$ and $\sigma_S$ are the water and the surface electrical conductivity (S m$^{-1}$), respectively.

As explained previously, the measured SP signal is a result of processes occurring within the geological medium ($SP^{EK}$ and $SP^{diff}$) and in the direct vicinity of the electrode (i.e., equilibrium electrode potential differences: $SP^{elec}$). Therefore, the measured SP signal at the $i^{th}$ electrode during the experiment can be obtained by the superposition of these signal components:

$$SP_i^{meas}(t) = SP_i^{EK}(t) + SP_i^{diff}(t) + SP_i^{elec}(t). \quad [27]$$

*2.3 Preprocessing and temperature filtering of SP data*

Differences in electrode temperature generate SP signals: $SP_i^T$ (section 2.1). For example, Maineult et al. (2004, 2005) and Allègre et al. (2010) display SP data with temperature effects, but did not attempt to correct them because of their relatively small amplitudes. Jardani et al. (2009) detrended their raw data to suppress an assumed linear signal attributed to temperature variations during a short time period (their Fig. 11). In other works, corrections are made using a relationship of the form,

$$SP_i^T = \alpha(T^i - T^{ref}). \quad [28]$$

where $\alpha$ is a sensitivity coefficient that depends on the electrode type: e.g. 0.2 mV/°C for Pb/PbCl$_2$, 0.7 or 0.9 mV/°C for Cu/CuSO$_4$, and -0.43 or -0.73 mV/°C for Ag/AgCl (Woodruff et al., 2010). This approach assumes a precise knowledge of the electrode temperatures over time, which implies that each electrode must be combined with a neighboring thermocouple. Even so, there is a risk that even small location differences between the electrodes and the thermocouples result in significant phase differences between the two locations.

Under laboratory conditions, the temperature variations in the vicinity of the electrodes are mainly dependent on the temperature distribution in the room and the heat transfer to the electrodes located within the medium (including the heat transfer through water displacement). Heat transfer in partially saturated porous media can be driven either by convection or conduction (e.g., Tabbagh et al. 1999). Thermal conduction is the most important process and it relies on two parameters that are strongly influenced by water saturation, namely, specific heat capacity and thermal conductivity. Even if it were possible to determine these parameters (Jougnot and Revil, 2010; Hamamoto et al. 2010), this would add additional complexity and uncertainties (the exact location of the electrodes with respect to the column must be known to minimize phase errors in the predicted diurnal variations) in the simulations. For this study, we choose instead to correct the SP data for temperature effects by applying a filter to each electrode dipole over pre-specified time periods (e.g., Coscia et al., 2012).

Assume first that we are interested in estimating a discrete and uniformly spaced time series of temperature at electrode $i$, $\mathbf{T}^i = \left(T_0^i, T_1^i, T_2^i, T_3^i, ..., T_N^i\right)^\tau$, with length $N$ given time series of ambient temperature in the laboratory defined as,

$$\mathbf{T}^{lab} = \left(...T_{-2}^{lab}, T_{-1}^{lab}, T_0^{lab}, T_1^{lab}, T_2^{lab},...\right)^\tau, \quad [29]$$

where the subscript indicates the time and superscript $\tau$ is the transpose operator. To do so, we postulate that there exists a linear filter $\mathbf{f}^{Ti}$ that if convolved with $\mathbf{T}^{lab}$ can predict $\mathbf{T}^i$ well,

$$\mathbf{T}^i \approx \mathbf{f}^{Ti} \otimes \mathbf{T}^{lab}. \quad [30]$$

This filter should be causal, which implies that it only comprises entries at positive times,

$$\mathbf{f}^{Ti} = \left(f_0^{Ti}, f_1^{Ti}, f_2^{Ti}, f_3^{Ti}, ... f_M^{Ti}\right)^\tau, \quad [31]$$

whith the length of the filter being $M$. In an analogous manner, we express the temperature vector $\mathbf{T}^{ref}$ in the vicinity of the reference electrode as: $\mathbf{T}^{ref} \approx \mathbf{f}^{Tref} \otimes \mathbf{T}^{lab}$. Our real interest resides in determining temperature difference over time, $\Delta \mathbf{T}^i = \mathbf{T}^i - \mathbf{T}^{ref}$, which given the linearity of the filters can be expressed as

$$\Delta \mathbf{T}^i \approx \mathbf{f}^{Ti} \otimes \mathbf{T}^{lab} - \mathbf{f}^{Tref} \otimes \mathbf{T}^{lab}. \quad [32]$$

We now combine Eq. [28] and [32] to express the predicted temperature effect on the SP time series at the $i^{th}$ electrode,

$$\mathbf{SP}_i^T = \alpha(\mathbf{f}^{Ti} - \mathbf{f}^{Tref}) \otimes \mathbf{T}^{lab}. \quad [33]$$

At the end of the drainage and imbibition cycles, when the previously described SP contributions (i.e., $SP^{EK}$, $SP^{diff}$, and $SP^{elec}$) are constant with time, the measured SP value between the $i^{th}$ and the reference electrode are only expected to vary due to temperature differences and a possible long-term linear electrode drift that is related to the aging of the electrodes. This gives in combination with Eq. [33],

$$\mathbf{SP}_i \approx a_i + b_i \mathbf{t} + \alpha (\mathbf{f}^{Ti} - \mathbf{f}^{Tref}) \otimes \mathbf{T}^{lab},  \quad [34]$$

where $a_i$ and $b_i$ are constants that quantify the SP signal offset and the drift, respectively, with $\mathbf{t}$ being the time vector. We introduce the SP-temperature filter function of electrode $i$ as $\mathbf{f}^i = \alpha(\mathbf{f}^{Ti} - \mathbf{f}^{Tref})$. For SP time-series that are significantly longer than the length of the SP-temperature filter functions, it is possible to determine its parameters through a linear inverse problem. Our model vector $\mathbf{m}_i$ for electrode $i$ combines $\mathbf{f}^i$, $a_i$ and $b_i$, while the design matrix $\mathbf{A}$ expresses the multiplication terms expressed in Eq. [34]. By solving the normal equation, the inversion minimizes the following cost function:

$$\min \Phi = \|\mathbf{Am} - \mathbf{SP}_i\|_2^2 + \lambda \|\mathbf{D}\|_2^2, \quad [35]$$

where $\lambda$ is a trade-off parameter between data fit and model smoothness and $\mathbf{D}$ is a roughness matrix that quantifies the gradient of $\mathbf{f}^i$. A large value of $\lambda$ will lead to a smoothly varying $\mathbf{f}^i$ that will possibly not allow explaining the observed data. To apply Eq. [35] it is important to determine periods during which the saturation is constant over time as the optimal filter is susceptible to change with saturation. The use of this filtering method necessitates a careful choice of the filter length $M$ and the trade-off parameter $\lambda$.

## 3. Material and methods

### 3.1 Choice of SP electrodes

Self-potential monitoring under laboratory conditions are mostly carried out using $Cu-CuSO_4$ or $Ag-AgCl$ electrodes. Electrodes of the first kind, such as $Cu-CuSO_4$, are usually immersed in a soluble salt of their cation $Cu^{2+}$ (e.g., Maineult et al., 2004), while electrodes of the second kind, such as $Ag-AgCl$, must be in contact with a solution containing the corresponding anion (see section 2.1). We based our choice of electrodes on tests by Tallgreen et al. (2005) who investigated the suitability of commercially available electrodes for slow electroencephalography (EEG) measurements. They found that the most stable signals are obtained by using sintered $Ag-AgCl$ electrodes immersed in a high chloride concentration gel (above $0.7$ mol $L^{-1}$). The sintered $Ag-AgCl$ electrodes are made by compaction and heating of the salt (AgCl) on the metal (Ag) to form a solid mass. This creates a layer of salt around the metal and avoids metal-anion interaction, which would generate a second electrode. We purchased the same electrode and gel brand as those that showed the best stability in this study.

The chosen electrode is a sintered $Ag-AgCl$ electrode from In Vivo Metric (E255). The electrodes consist of an Ag wire with a sintered solid AgCl salt along 0.7 cm of the wire. Plastic insulation leaves only a small cylindrical part as a contact (0.1 cm-diameter and 0.25 cm-long). We also insulated the Ag wire from light using a thermo-retractable sheath to avoid photoelectric effects. Then, we constructed PMMA (polymethyl methacrylate plastic) chambers ($0.716$ $cm^3$) and filled them with a high concentration chloride gel (close to $0.7$ mol $L^{-1}$ according to Tallgreen et al. (2005)). Spherical shape porous ceramics were pasted to the chambers and their contact surface area with the geologic medium was reduced to $0.07$ $cm^2$ using epoxy resin to minimize gel leakage, but large enough to allow an electrical contact with a porous medium (tests were performed using sands at different saturations). In previous studies, chambers with porous ceramics often have larger contact surface areas: for example, around 1.7 and 1.5 $cm^2$ in Allègre et al. (2010) and Mboh et al. (2012), respectively, while the porous ceramic used herein has the same thickness (0.2 cm) as in these works.

Prior to any SP measurements, we designed a simple experimental set-up to quantify electrode leakage of our home-designed electrode (In Vivo Metric Ag-AgCl electrode immersed in a gel filled chamber). Figure 3a presents a sketch of the experimental set-up (not to scale). The electrode was immersed in a reservoir filled with deionized water (60 cm$^3$) that was sealed to avoid evaporation. The initial electrical conductivity was $\sigma_w = 6$ μS cm$^{-1}$ and its evolution over time was monitored by an immersed conductometer at a fixed position. The water was constantly homogenized using a magnetic stirrer to avoid local concentration gradients. A second test was conducted with a widely used SP field electrode (SDEC NaCl Petiau electrode). The Petiau (2000) electrode is made of a 18 cm-long plastic tube, forming a chamber, which is filled by a clay mud saturated by a fairly saline pore water with an acidic pH (between 4 and 5) to ensure a high degree of chemical stability of the electrode. The contact with the geological medium is achieved through a wood plug (~6.15 cm$^2$).

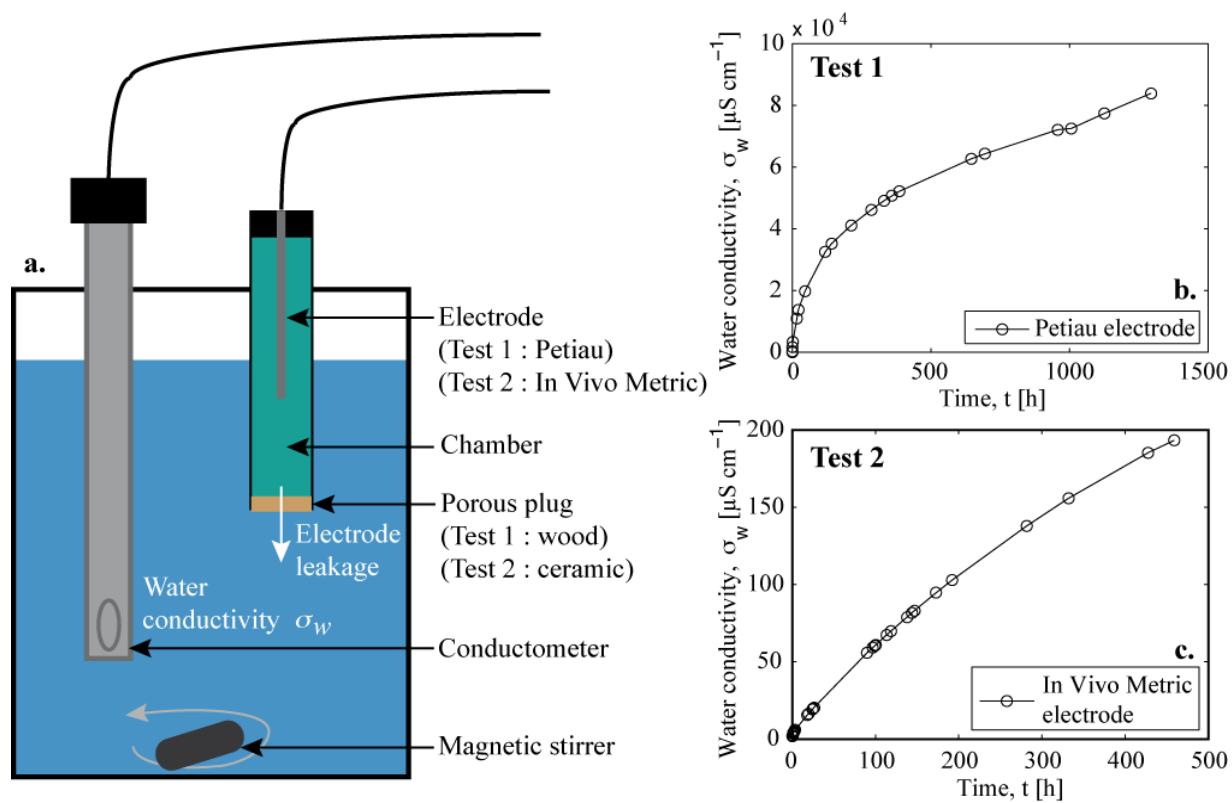

*Fig. 3. Electrode chamber leakage test: (a) sketch of the experimental set-up, (b) and (c) water conductivity reservoir evolution with time for Test 1 (Petiau electrode) and Test 2 (In Vivo Metric electrode within a gel filled chamber), respectively.*

The leakage experiment with the Petiau electrode (stopped after 1400 hours) showed a very important increase in electrical conductivity at the end of the experiment (close to $9 \times 10^4$ μS cm$^{-1}$, Fig. 3b). The test with the Ag-AgCl electrode was stopped after 500 hours and showed a strong increase of the water conductivity (200 μS cm$^{-1}$, Fig. 3c). Even after several hundreds of hours, the electrodes are still leaking due to persistent concentration gradients between the chamber and the water reservoir. The differences in conductivity between the two reservoirs at the end of the experiments can be explained by differences in the contact surface (two orders of magnitude), the initial chamber chemistry, and the duration of the experiments.

The simple experiments described above illustrate that leakage from an electrode chamber cannot be ignored for longer-term experiments in the laboratory. It appears impossible to use the very stable Petiau electrodes for laboratory purposes without severely polluting the studied medium. The In Vivo Metric electrode with the gel filled chamber show a smaller leakage, which can possibly be ignored for short-term measurements, but not for monitoring periods involving hours or days. Another complication is that diffusion coefficients are saturation dependent (e.g., Revil and Jougnot, 2008), which leads to different leakage rates. Considering all these issues, we decided to conduct our monitoring of drainage/imbibition cycles without any chambers, that is, to insert the Ag-AgCl electrodes directly in the geological medium (see Linde et al., 2007 for a similar set-up). In this configuration, the electrode surface described in Fig. 1 is in direct contact with the pore water.

*3.2 Experimental drainage/imbibition cycles*

We monitored SP during cycles of drainage and imbibition experiments in a PMMA column filled with a rather homogeneous sand, namely the Fisher scientific sand used by Linde et al. (2007). This material is mainly made up of silica: $SiO_2$ (>95%), $KSi_3AlO_8$ (4%), and $NaAlSiO_8$ (<1%). The grain diameter is comprised between 100 and 160 μm, with a mean diameter of 132 μm. Before packing the experimental column, we immersed the sand in an electrolyte bath. We used a binary symmetric [1:1] NaCl electrolyte with a chloride concentration $C_{Cl^-} = 1.9 \times 10^{-3}$ mol L$^{-1}$ and an electrical conductivity $\sigma_w = 200$ μS cm$^{-1}$. The sand was left during two weeks in the water bath to equilibrate. We renewed the water every week (i.e., two times) as the water conductivity tended to increase over time (up to 400 μS cm$^{-1}$, see also appendix A of Allègre et al. (2010)). The column has an internal diameter of 5.92 cm and a total height of 165 cm, but it was filled with sand up to 150 cm. We considered the bottom of the column to be the elevation reference (i.e., $z = 0$ cm). To keep the sand saturated, we took it directly from the bath in doses of 50 g saturated sand until the column was filled. We always kept no more than 1 cm of the equilibrated water above the sand to avoid sedimentation effects, entrapped air and to control the packing heterogeneity in the column. We let the grains reorganize themselves by natural packing in the column for 2 days before re-adjusting the sand level to 150 cm. The final porosity in the column was $\phi = 0.427$.

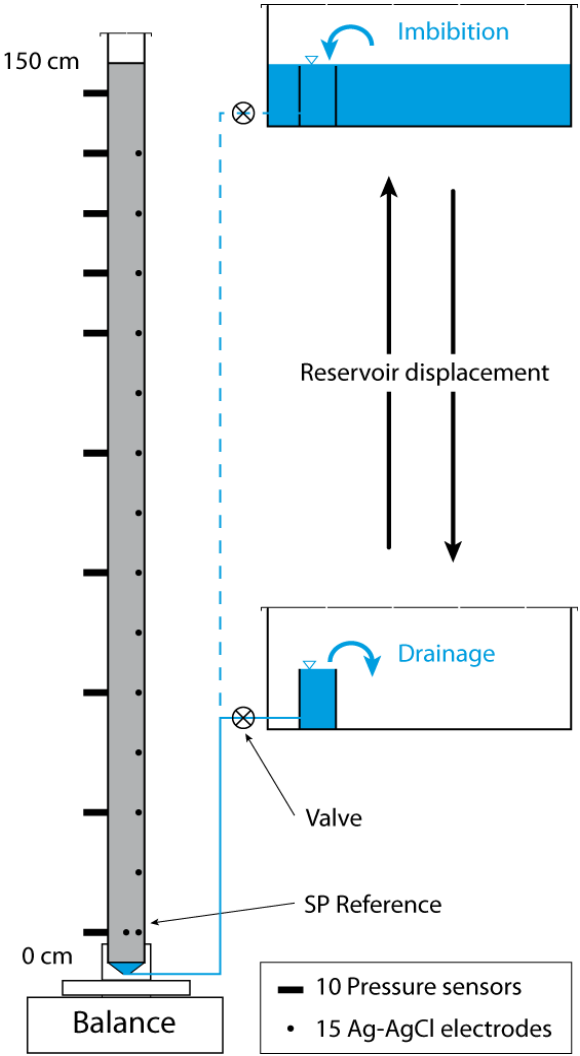

*Fig. 4. Sketch of the experimental set-up for the drainage-imbibition cycles.*

Figure 4 presents the experimental set-up. The bottom of the column was hydraulically connected to an external reservoir by a PTFE (polytetrafluoroethylene) tube with a 0.2 cm internal diameter; the total volume of the connection between the bottom of the sand and the reservoir is 83.2 cm$^3$. The elevation of the reservoir imposed a pressure at the bottom of the column. The reservoir was left within a bigger one that could retain the over flow of draining water without changing the hydraulic pressure in the column. The second reservoir was also chosen sufficiently large to impose a quasi-constant hydraulic pressure during imbibition; the water volume entering the column corresponds to less than a 0.5 cm height in the larger reservoir (i.e., less than the uncertainty of the pressure sensor). A valve was installed to reduce pressure perturbations while moving the reservoir. After moving the reservoir, the valve was closed for 3 to 5 hours before starting a new hydrological event (drainage or imbibition). Before initializing the experiments, the water level in the column and the reservoir were maintained above the sand level for three days to ensure the absence of water movement prior to any imposed hydrological events. The top parts of the column and reservoirs were covered to avoid (or at least limit) evaporation, while small holes were made in the cover to allow for pressure changes.

The column installation was placed on a balance and equipped to monitor water pressure and SP at different elevations. The balance (Mettler Toledo, MS32001LE) can measure up to 32 kg with an accuracy of 0.1 g. It was connected to a computer and the column system mass was stored every 5 s. The pressure was monitored with 10 pressure sensors (Sensor Technics, CTE8N01GY7), which can measure between -1 and +1 bar (i.e., ± 10 m of water) with an accuracy of ± 0.2 % full-scaled output. They are located at $z$ = 5, 25, 45, 65, 85, 105, 115, 125, 135, and 145 cm. The pressure sensors are connected to water filled chambers that are in contact with the medium through a porous ceramic to allow measuring the water phase pressure under partially saturated conditions. These ceramics have a half spherical cylindrical shape and a 0.5 cm diameter. Self-potential monitoring was conducted with Ag-AgCl electrodes from In Vivo Metric inserted ~1 cm into the sand at 15 locations: every 10 cm, from $z$ = 5 to 135 cm. Two electrodes were installed at the bottom of the column ($z$ = 5 cm), one was chosen as the reference electrode during the measurements. Pressure and SP signals were scanned every second by a data logger CR3000 (Campbell Scientific) with an internal resistance $R$ = 20 GΩ. The signals were filtered to remove the 50 Hz signal from the electrical network, and values averaged over 5 s were stored in the data logger. In addition, the air temperature $T_{air}$ (°C) and relative humidity $RH$ (%) in the laboratory (30 cm away from the column) were stored every 20 minutes. Before and after each hydrological event, the temperature $T_w$ and the electrical conductivity $\sigma_w$ were measured in the small reservoir.

Due to the 2 day stabilization period before initializing the measurements, we found after the first drainage that the electrical conductivity of the outflux water had increased to $\sigma_w$ = 360 μS cm$^{-1}$. To avoid an electrical conductivity contrast in the pore water of the column that could complicate the interpretation of the SP measurements, we completed the filling of the bigger reservoir with a NaCl solution with the same electrical conductivity (but a higher chloride concentration than the water in place, $C_{Cl^-} = 2.4 \times 10^{-3}$ mol L$^{-1}$ for the imbibition water entering the column). This results in a difference in chloride concentration between the water in the column and in the reservoir. We conducted 19 cycles of drainage and imbibition starting at different elevations and with different amplitudes during 6 months. In the present study, we focus on the SP monitoring of the first two drainage/imbibition cycles. For both cycles, the reservoir was displaced from $z = 150$ cm to 49 cm before it was moved back to 150 cm. The datalogger was turned on 4.5 h before starting the experiments. The beginning of drainage 1 is used as initial state ($t = 0$ h), imbibition 1 was initialized after 72.78 h, drainage 2 after 144.28 h and imbibition 2 after 216.59 h.

### *3.3 Hydrodynamic and electrical characterization of the sample*

Additional hydraulic and electrical characterization of the sand was needed to conduct the numerical simulations. The hydraulic and electrical parameters of the sand are provided in Tables 1 and 2, respectively.

*Table 1: Hydrodynamic parameters of the sand used in the experiments*

|  | $\phi$ (-) | $\theta_w^r$ | $\alpha_{VG}$ (Pa$^{-1}$) | $m_{VG}$ (-) | $k$ (m$^{-2}$) |
|---|---|---|---|---|---|
| Linde et al. (2007) | 0.33-0.35 | - | $1.54 \times 10^{-4}$ [a.] | $0.90$ [a.]/$0.87$ [b.] | $7.9 \times 10^{-12}$ |
| Sample measurements | 0.403 [c.] | 0.011 [c.] | $1.84 \times 10^{-4}$ [c.] | 0.858 [c.] | $1.54 \times 10^{-11}$ [d.] |
| Drainage 1 | 0.427 | 0.143 [e.] | $2.16 \times 10^{-4}$ [e.] | 0.904 [e.] | $1.53 \times 10^{-11}$ [e.] |
| Imbibition 2 | 0.427 | 0.106 [e.] | $1.98 \times 10^{-4}$ [e.] | 0.876 [e.] | $1.61 \times 10^{-11}$ [e.] |

  a. from inversion for the water retention function (Eq. [12])
  b. from inversion for the relative permeability function (Eq. [14])
  c. from suction table measurements
  d. from falling head measurements
  e. from HYDRUS1D inversion

In the column used by Linde et al. (2007), the sand had a lower porosity $\phi = 0.33 - 0.35$, which can be explained by a different packing method, and hence a lower permeability $k = 7.9 \times 10^{-12}$ m$^2$. The authors obtained van Genuchten parameters $\alpha_{VG}$ and $m_{VG}$ through inversion of the capillary pressure and outflow drainage data. Considering the electrical parameters of the sand, they used a formation factor, *F*, and a saturation exponent, *n*, from the literature (Table 2). The authors also determined the saturated effective excess charge of the medium $\bar{Q}_v^{eff,sat} = 0.48$ C m$^{-3}$ from experiments under saturated conditions.

The water retention function was characterized using suction tables. Although the porosity of the sample is higher, the van Genuchten parameters obtained by these measurements are in good agreement with the inversion results of Linde et al. (2007). The hydraulic conductivities of five samples were measured by the falling head technique yielding a higher permeability ($k = 1.54 \, (\pm \, 0.03) \times 10^{-11}$ m$^2$) than by Linde et al. (2007).

*Table 2: Electrical parameters of the sand used in the experiments*

|  | $\sigma_w$ (µS cm$^{-1}$) | $F$ (-) | $m$ (-) | $n$ (-) | $Q_v^{eff,sat}$ (C m$^{-3}$) |
|---|---|---|---|---|---|
| Linde et al. (2007) | 510 | 4.24 | 1.34 | 1.6 | 0.48 [a.] |
| This study | 360 | 3.27 [b.] | 1.40 [b.] | 1.6 | 0.47 [c.] |

    a. inverted from SP measurements

    b. measured by spectral induced polarization

    c. computed by the Jougnot et al. (2012) model using the water retention function derived from inversion of drainage 1 (Table 1)

The electrical properties of the sand were measured under saturated conditions using the spectral induced polarization device developed by Zimmermann et al. (2008). Measurements were operated within the 10 mHz to 45 kHz frequency range with the same electrodes as Koch et al. (2012). Seven sand samples were left to equilibrate in NaCl solutions with different salinities ranging from 30 to 2500 µS cm$^{-1}$. Due to the increase of pore water salinity over time (see previous section), the lowest salinity solutions were replaced at several occasions before the measurements. The electrical conductivity measurements for each sample were repeated several times and the pore water conductivity was checked before and after the measurements. Figure 5 shows the results of these measurements and the best fit of Eq. [26]. We find that the surface conductivity of the sand can be safely neglected. The formation factor is smaller than in the study of Linde et al. (2007) because of the higher porosity in the present study.

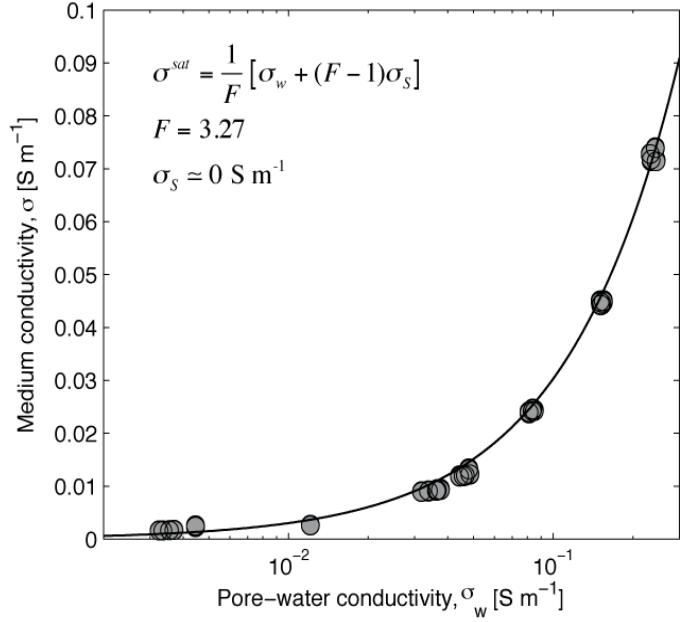

*Fig. 5. Electrical conductivity measurements of the sand at different pore water salinity.*

The hydrodynamic parameters of the sand in the packed column were obtained from the capillary pressure data from the first two cycles using Levenberg–Marquardt optimization in HYDRUS1D (Šimůnek et al., 2005). Considering the porosity to be known, we determined four parameters from the pressure data assuming that the values were constant within the column: $\theta_w^r$, $\alpha_{VG}$, $m_{VG}$, and $k$ (Table 1). The previously obtained parameter values were chosen as the initial model for the inversion algorithm. The drainage and imbibition events were inversed independently. The obtained parameter values are in fairly good agreement with the sample measurements (Table 1).

The effective excess charge function $\bar{Q}_v^{eff}(S_w)$ was determined using the method proposed by Jougnot et al. (2012). We computed it from the estimated hydrodynamic parameters from both drainage 1 and imbibition 2 (Table 1) and the water retention (WR) approach described in this paper. We choose not to use the second approach proposed by Jougnot et al. (2012), which is based on the relative permeability (RP) because in our case this function was not well constrained (i.e., no direct measurements available). The effective excess charge value at saturation are $\bar{Q}_v^{eff,sat}$ = 0.47 and 0.55 C m$^{-3}$ obtained from drainage 1 and imbibition 2. Figure 6 shows the resulting relative effective excess charge function $\bar{Q}_v^{eff,rel}(S_w)$ together with the Linde et al. (2007) model, in which the excess charge scales with the inverse of the saturation:

$\bar{Q}_v^{eff}(S_w) = \bar{Q}_v^{eff,sat}/S_w$. The effective excess charge function obtained with the water retention approach give results that are fairly similar to the model of Linde et al. (2007). The empirical relationship by Jardani et al. (2007) that relates permeability and effective excess charge predicts $\bar{Q}_v^{eff,sat} = 0.44$ C m$^{-3}$, which is fairly close to the predicted values from the water retention approach (see Table 2). In a recent paper, Mboh et al. (2012) showed that the Linde et al. (2007) model works for a well-sorted quartz sand for the saturations considered herein. Jougnot et al. (2012) demonstrate that such a good agreement is not observed for many natural porous media for which pore size distributions are much wider (e.g., soils, dolomite).

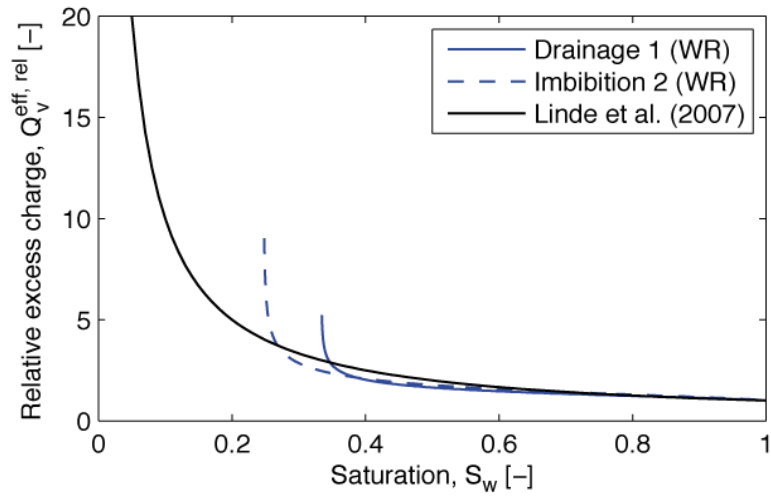

*Fig. 6. Comparison between the four $\bar{Q}_v^{eff,rel}(S_w)$ functions obtained from the hydrodynamic parameters following the WR approach of Jougnot et al. (2012). The model of Linde et al. (2007) is shown for comparison.*

We compared the simulation results based on the hydrodynamic parameters obtained from the inversion of the drainage 1 and imbibition 2 data to the pressure data in the column during the full observational period, and found that both sets of parameters could reproduce the observations within the measurement precision. We tested the two $\bar{Q}_v^{eff}(S_w)$ functions obtained from the Jougnot et al. (2012) WR approach for these simulations. The SP amplitudes were better reproduced by the values obtained from drainage 1 and we use those in the following.

The solute transport simulations were performed using the water velocities obtained with the chosen set of parameters and the NaCl molecular diffusion coefficient in a dilute solution $D_{NaCl}^w = 1.6 \times 10^{-11}$ m$^2$ s$^{-1}$. The vertical dispersivity was adjusted to obtain the best match to the SP data, which resulted in $\alpha_v = 10^{-2}$ m. In the simulation, the Cl$^-$ concentration of the water entering the column for imbibition 1 and 2 was adjusted according to the experimental procedure ($C_{Cl^-} = 2.4 \times 10^{-3}$ mol L$^{-1}$). In the modeling, we also took the volume of water contained in the tube with the initial chloride concentration $C_{Cl^-} = 1.9 \times 10^{-3}$ mol L$^{-1}$ into account (see section 3.2). As sodium and chloride are the main ionic species in the pore water, we only consider them for the electro-diffusion current generation. The microscopic Hittorf number of these two ions calculated from their ionic mobilities, $\beta_{Cl^-} = 5.19 \times 10^{-8}$ m$^2$ s$^{-1}$ V$^{-1}$ and $\beta_{Na^+} = 8.47 \times 10^{-8}$ m$^2$ s$^{-1}$ V$^{-1}$, yield using Eq. [22]: $t_{Cl^-}^H = 0.620$ and $t_{Na^+}^H = 0.380$. This is a simplification of the system, given that the exact pore water chemistry is unknown (i.e., ions dissolved in the pore water during the stabilization period). In section 2.2, we assumed that the electrical double layer would not strongly affect the diffusion in the pore space. This is confirmed by considering that the diffuse layer length is estimated to be $1.4 \times 10^{-8}$ m, while the characteristic pore throat size, also called Johnson length, is approximated to $1.4 \times 10^{-5}$ m (e.g., Revil et al., 1999). We simulated the electrical problem using the parameters obtained from our measurements; in addition, we used a literature value for the Archie's saturation exponent (Table 2). Note that the electrical conductivities are calculated for an assumed constant $\sigma_w = 360$ μS cm$^{-1}$.

The theoretical framework described in section 2.2, together with the medium parameters (Table 1 and 2), was implemented in the finite element code COMSOL Multiphysics 3.5. The experimental column is represented by a numerical domain with a length of 150 cm with measurement points corresponding to the sensor locations in the experimental set up (see section 3.2). This geometry is discretized into more than 600 elements with different sizes depending on the expected saturation during the experiment: smaller than 1 cm between the bottom and a height of 75 cm, smaller than 0.5 cm at 75 – 125 cm, and smaller then 0.25 cm between 125 cm and the top of the column. Drainage and imbibition cycles are initiated and driven by imposed pressure changes at the bottom of the column. The chemistry of the water entering the column is controlled by a time dependent boundary condition. The mass leaving or entering the system is calculated from the water flows across the lower boundary and the incoming or outgoing concentrations. Neumann boundary conditions are imposed to the lateral and upper sides of the column for the flow and transport problem (i.e., no flux). The system is considered to be in hydrostatic equilibrium prior to any imposed change. For the electrical problem, Neumann boundary conditions (i.e., electrical insulation) are also applied to all sides and the electrical reference voltage ($\varphi = 0$ V) is imposed at the same location as in the experiment ($z = 5$ cm). Flow, transport, and source current generation are simulated by solving the three corresponding equation systems (Eq. [13], [16], and [24]) with a segregated time dependent solver.

## 4. Results

### *4.1 Monitoring data and temperature filtering*

Figure 7 shows the raw data from the first two monitoring cycles. For the SP data (Fig. 7d), we use the second bottom electrode ($z = 5$ cm) as a reference and we set all the SP data to zero prior to the first drainage.

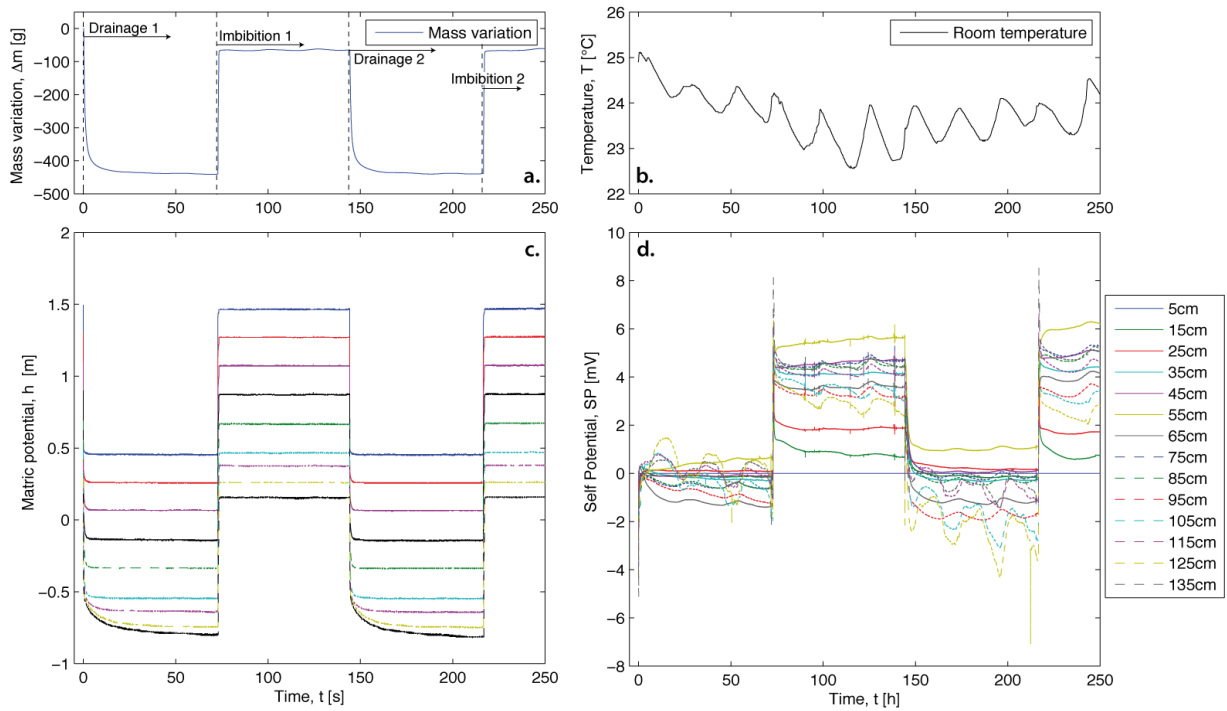

*Fig. 7. Monitoring results for the first two drainage/imbibition cycles: (a) mass variation in the column from the initial conditions, (b) the room temperature, (c) the matric potential and (d) the raw SP data.*

The two drainage and imbibition events can clearly be observed in the mass variation of the column (Fig. 7a), in the pressure data (Fig. 7c), and in the SP data (Fig. 7d), but the SP data also display significant diurnal oscillations that are unrelated to the pressure changes. The pressure change in the medium responded to the four reservoir displacements: ~100 cm decrease at 0 and 144.28 h, and corresponding increases at 72.78 and 216.6 h. The SP values show a negative peak when drainage 1 starts (0 h) as expected (Linde et al., 2007; Mboh et al., 2012). Two positive peaks are related to imbibitions 1 and 2 (72.78 and 216.59 h) as expected, but never shown in published experimental data. After imbibition 1, the SP signal did not relax to zero even though the hydraulic conditions were almost equivalent to the initial state (saturated and no water flux).

The changes in room temperature during the experiment were rather small (2.5°C; Fig. 7b), but their effects on the SP data are non-negligible. The measured temperature effects are of similar amplitude as those measured by Allègre et al. (2010), but no temperature effects can be seen in the data of Mboh et al. (2012). It is possible that the chamber electrodes used by Mboh et al. (2012) help to have more synchronous changes in temperature around the electrodes and therefore minimize the temperature effects on the SP measurements. It appears that the temperature effects on the SP data are more important for the electrodes in the unsaturated part of the column (see 75 cm and above, after drainage 1 and 2). For example the diurnal SP variations at 125 cm is around 2 mV during or after the drainage (Fig. 8a and c), but less than 1 mV after the imbibition (saturated condition, Fig. 8b and d). This observation indicates that the saturation has an impact upon the temperature difference (magnitude and phase) between the reference and the $i^{th}$ electrode, due to differences in thermal transfer in the porous medium as predicted by theory (e.g., Jougnot and Revil, 2010).

We corrected the data for temperature effects through the filtering method described in section 2.3. We considered the four stabilization periods following the four hydrological events and determined the representative electrode-specific filters for those time periods using periods during which no changes in mass were observed. Figure 8 presents the filtering results for the first two stabilization periods. We performed a sensitivity study by varying the filter length from 2 to 12 h and the trade-off parameter $\lambda \in [5 \times 10^{-5}; 0.5]$. We chose a $M$ corresponding to 6 h, which was long enough to adequately remove most of the temperature effects. Given that the room temperature variations are mainly diurnal and that we are interested in the phase lag between temperature variation at the $i^{th}$ and the reference electrode, a quarter of the dominant period (i.e., 6 h) appears to be a good choice. We chose $\lambda = 0.01$ because it was the largest value (i.e., the smoothest model) for which we could adequately correct the data. The estimated filters display the largest amplitudes at early times (Fig. 8e and f), representing short time delays in temperature between the electrodes, and it is clear that the temperature effect is the most important during unsaturated conditions (i.e., after drainage, Fig. 8g). The filters determined for the electrodes in the nearly saturated medium (75 to 95 cm in Fig. 8e) are fairly similar to the ones in the fully saturated medium (same electrodes in Fig. 8f). Note that the filter only correct the signal for temperature effects (e.g., visible drifts remain in the filtered data, Fig. 8i and j). The filtering process gives satisfying results: for example, the biggest SP variation due to the temperature effect (2 mV for electrode at 125 cm on Fig. 8c) is reduced to less than 0.5 mV (Fig. 8g). The filters were then applied to cover the full time series (i.e., from the beginning of an event to the next one). The result of this filtering is shown in Fig. 9c, illustrating that most of the temperature perturbations have been filtered out.

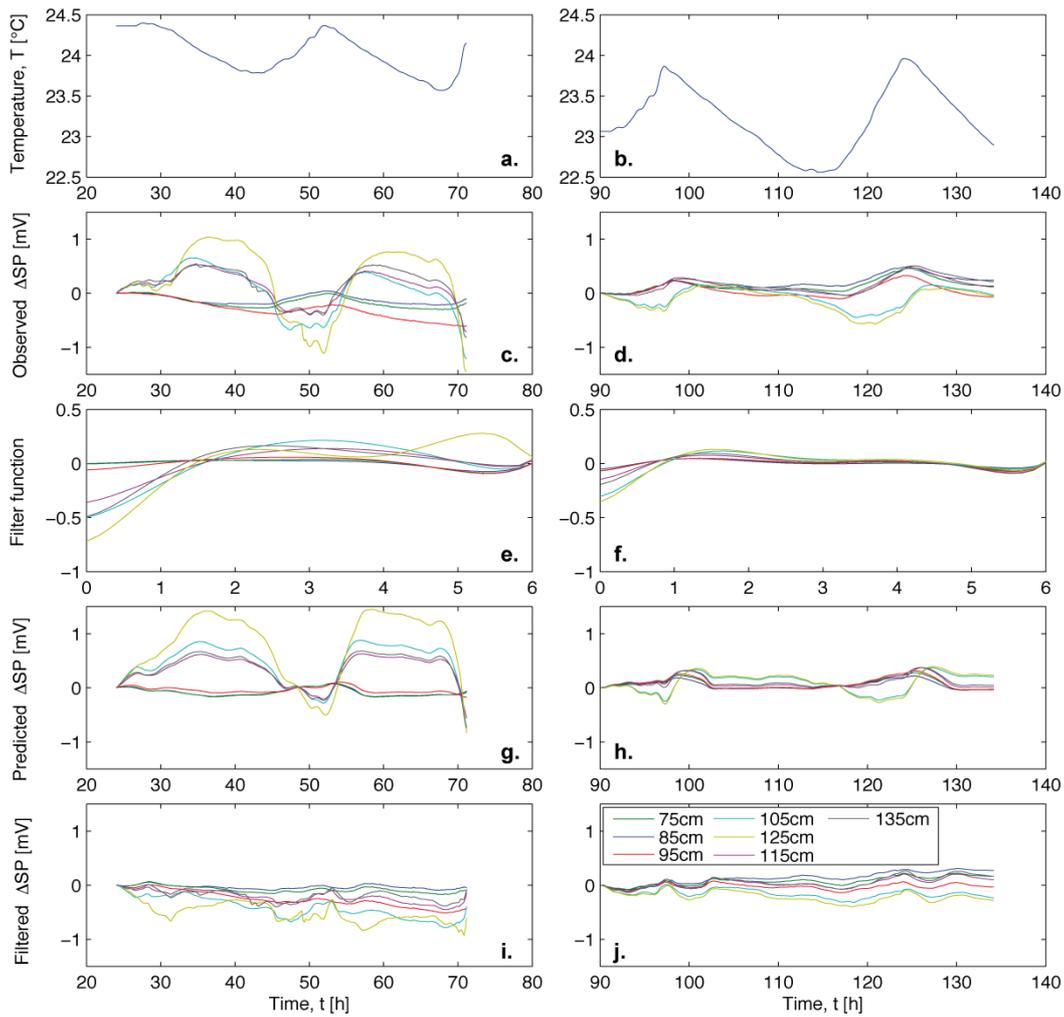

*Fig. 8. Temperature filtering after drainage 1 (partially saturated medium, left column) and imbibition 1 (fully saturated medium, right column): (a. and b.) measured room temperature, (c. and d.) observed SP variations, (e. and f) determined filter functions, (g. and h.) predicted temperature effect, and (i. and j.) temperature filtered SP variation. Only the upper electrodes for which the temperature effects are the most important are shown.*

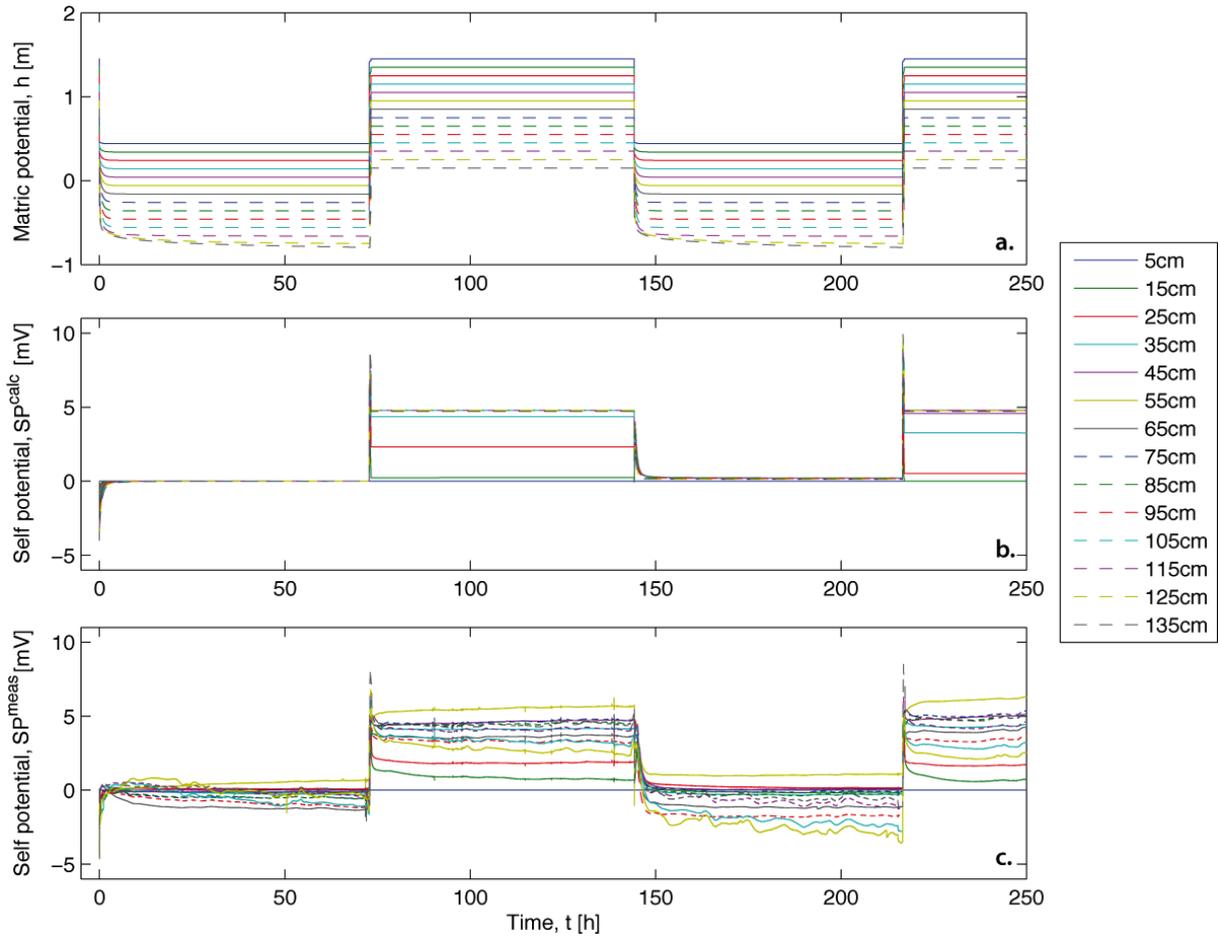

*Fig. 9. Results of the numerical simulation of (a) the matric potential and (b) the SP, and (c) the temperature-filtered SP measurements for the first two drainage and imbibition cycles.*

Figure 9a shows simulated pressure data for the first two cycles. The simulation reproduces the observed pressure measurements within 2 cm, except for the upper three pressure sensors, which were reproduced within 5 cm (Fig. 7c). Figure 9b shows the numerical simulation of the SP data for the same time period. The main features of the temperature-filtered SP data (Fig 9c; referred to as measured data in the following) are well captured by the numerical model. The first drainage peak reaches -4.3 mV (at 0 h) and is followed by a short relaxation time of ~6 h, which is similar to previous studies (e.g., Linde et al., 2007; Mboh et al., 2012). Both the simulated and measured data show a positive SP peak close to +10 mV in response to imbibition 1 (at 72.78 h). This event is followed by a relaxation, in which the signals do not return to zero. In the simulation, the bottom electrodes (5, 15, and 25 cm) equilibrate gradually between 0 and 4.8 mV (0.2, 2.4, and 4.5 mV, respectively), while the other electrodes equilibrate around 4.8 mV despite that no water flow occurs and the column is fully saturated. This residual signal is attributed to the distribution of chloride concentration along the column at the end of the imbibition. In response to drainage 2, the expected negative peak is masked by a gradual decrease of the SP values from the previous stabilization ($SP_i^{elec}$) to 0 mV for the simulation (Fig. 9b) and to a set of values scattered around 0 mV in the measurements. Finally, the SP response to imbibition 2 (at 216.59 h) is fairly similar to imbibition 1, with a positive peak of the same amplitude.

*4.2 Decomposition of SP contributions*

In the previous section, we showed that our numerical simulation was able to capture and reproduce the main features of the temperature-detrended SP monitoring data during two drainage/imbibition cycles. To investigate the influence of each contribution to these data, we here focus on the hydrological events on a shorter time scale (2 h). Each SP response to an event is decomposed into the three simulated signal components (Eq. [27]): the electrokinetic ($SP^{EK}$), electro-diffusive ($SP^{diff}$), and the equilibrium electrode potential differences ($SP^{elec}$). The sum of these polarizations ($SP^{calc}$) can then be compared to the measured data ($SP^{meas}$). Figures 10 and 11 regroup the two drainage and imbibition periods, respectively. To facilitate comparison, we set the SP magnitudes to zero before each corresponding event. We also shifted the time scale, such that $t = 0$ h corresponds to the beginning of the considered hydrological event (drainage or imbibition).

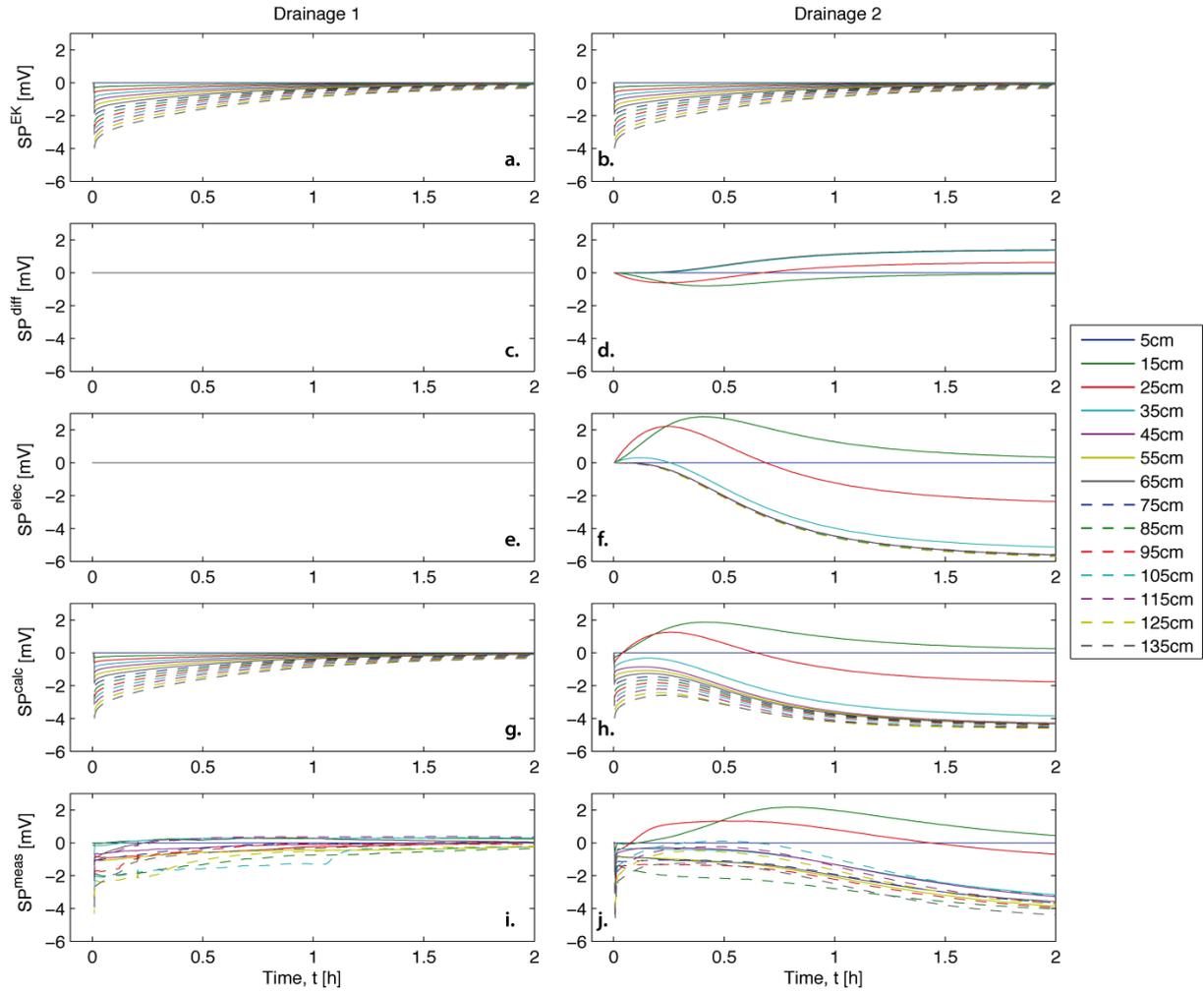

*Fig. 10. The simulated contributions of the SP signal in terms of (a, b) electrokinetic, (c, d) electro-diffusive, and (e, f) equilibrium electrode potential difference, (g, h) the total calculated and (i, j) measured (temperature-detrended) SP during drainage 1 and 2.*

The electrokinetic contributions simulated in response to the two drainage events (Fig. 10a and b) are similar. The simulated electro-diffusive polarization and electrode potential differences during drainage 1 are zero because the sodium and chloride concentration distribution is assumed to be homogeneous in the medium (initial pore water), while this is not the case when starting drainage 2 (Fig. 10c and e). In fact, the imbition water, which entered the column during imbibition 1 had a chloride concentration that was 1.26 times higher than the initial pore water, even if the electrical conductivity was the same (see sections 3.2 and 3.3 for experimental and simulation details, respectively). This created equilibrium electrode potential differences that are directly linked to the Cl$^-$ concentration in the vicinity of the electrodes (Fig. 1). This electrode effect has been calculated using Eq. [10] and the chloride transport simulation results and Figure 10f clearly shows that it affected all electrodes as the $C_{Cl^-}$ changes occurred principally in the bottom of the column, thus in the vicinity of the reference electrode. The resulting magnitude (up to 6 mV) of this electrode effect is more important than the electrokinetic one (up to 4 mV). In the mixing zone between the imbition pore water and the water in place, gradients of chloride and sodium concentration prevail. The electro-diffusive polarization due to NaCl diffusion given by the right hand side of Eq. [25] is fairly small (less than 1.5 mV; Fig 10d) in comparison to the equilibrium electrode potential difference. The phenomena related to changes in NaCl concentration result in different responses in the total calculated signal, $SP^{calc}$, for the two drainage events (Fig. 10g and h) and thereby illustrate the importance of considering electrode effects when interpreting SP measurements. For the first drainage, the correspondence between the simulated responses and the data is satisfying in terms of amplitude (a peak around -4.3 mV) and relaxation, although the observed signals did not relax to 0 mV as predicted by the model. Drainage 2 is fairly well reproduced in terms of the peak amplitude (-4.3 mV) when the drainage starts and the following complex behavior at the different electrodes. In drainage 2, the influence of the electrode potential difference is basically driving the entire signal trend, explaining the increasingly negative values of the upper electrodes.

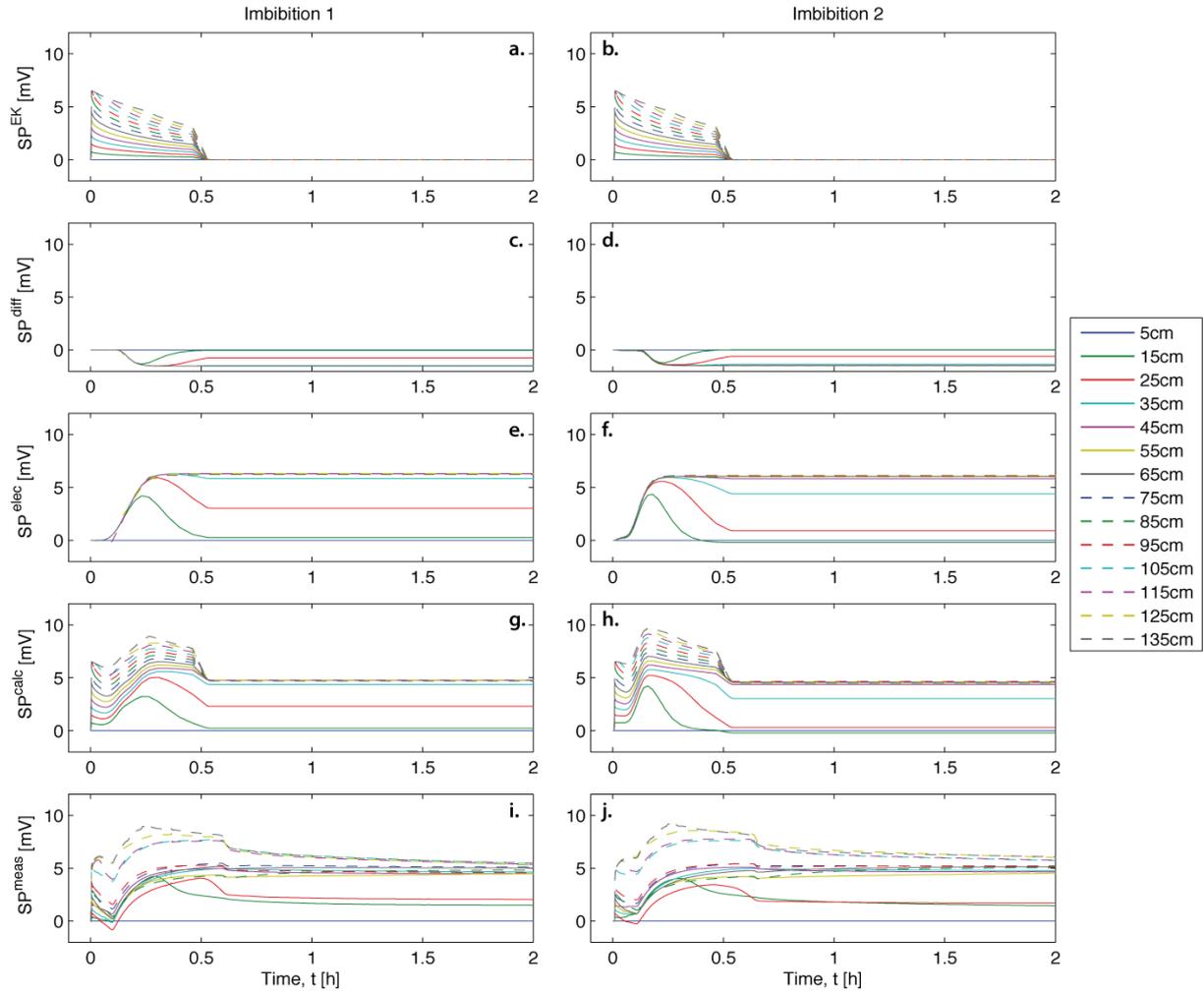

*Fig. 11. The simulated contributions of the SP signal in terms of (a, b) electrokinetic, (c, d) electro-diffusive, and (e, f) equilibrium electrode potential difference, (g, h) the total calculated and (i, j) measured (temperature-detrended) SP during imbibition 1 and 2.*

Unlike the two drainages, the SP signals for imbibitions 1 and 2 are fairly similar (Fig. 11). The simulated electrokinetic responses to the imbibitions show a peak of up to 6 mV followed by relaxations, which stop abruptly ~0.55 h after the start of the imbibition (Fig. 11a and b). This sharp transition corresponds to the end of the imbibition when the mass and pressures stabilized. The electro-diffusive contribution to the SP (Fig. 11c and d) and electrode effects (i.e., due to chloride concentration differences) started between 3 to 7 min after the imbibition because of the time needed for the imbibition water (with a higher sodium and chloride concentration) to reach the reference electrode and thereby generate these two signal contributions (Fig. 11e and f). The electro-diffusive contributions have a reverse sign and smaller amplitude (up to 1.5 mV) compared with the electrode effects (up to 6.5 mV). When the imbibition ends, these two signals stabilize and generate a residual signal that is directly linked to the distribution of the sodium and chloride concentrations in the column. $SP^{EK}$ and $SP^{elec}$ have the same amplitudes for the upper electrodes. As the electrode effects occurred with a delay, the sum of the two contributions presents a bimodal shape. The total simulated self-potential, $SP^{calc}$ displays a first peak (up to 6 mV) when the imbibition starts, followed by a short relaxation interrupted by a new increase up to 10.2 mV. When the imbibitions stop, the residual values are mainly due to the chloride concentration distribution in the column (Fig. 9c). The simulated and the measured SP data are in fairly good agreement in terms of the characteristic shapes and amplitudes. The experimental data for the two imbibitions are similar (Fig. 11i and j). They show a first peak (up to 5 mV) followed by short relaxation and a larger and smoother second peak (up to 12 mV). The data also show some breaks in the curves, which correspond to the end of the imbibition. The actual imbibition appears to have lasted somewhat longer than in the simulations, which can be explained by smaller errors in the parameterization of the flow model. The discrepancies between the measured and the simulated amplitudes are mainly attributed to local scale heterogeneities in the chloride concentration that are unaccounted for in the transport model.

## 5. Discussion

*5.1 Self-potentials as a multi-process sensor*

Our results clearly show that, even in well controlled laboratory conditions, the SP method senses different processes at the same time and that each possible contribution to the SP signal, arising either from the medium or the electrodes, has to be considered prior to any quantitative interpretation of the data in terms of one preferred mechanism. The main interest in the SP method for vadose zone hydrologists is traditionally linked to the electrokinetic source (generating the so-called streaming potential, $SP^{EK}$). This process is rather well understood (Linde et al., 2007; Jougnot et al., 2012), but the results presented herein clearly illustrate that this is only one of the components of the total measured SP and that other contributions must be carefully considered. These are: temperature differences, electro-diffusive processes, and equilibrium electrode potential differences.

From section 2, it is evident that temperature affects many properties, such as the electrode standard potential (Fig. 2), the equilibrium electrode potential (Eq. [9]), the electro-diffusion (Eq. [23]) and also the medium electrical conductivity. Figure 7 shows that, even small diurnal variations of the room temperature (1°C) may yield SP signals of a few mV (i.e., similar in magnitude to $SP^{EK}$). Given experimental uncertainties, it might be hazardous to try to model all these processes related to temperature. In this study, we filtered out most of the temperature effects from the SP signal through a deconvolution procedure, in which we capitalized on periods during which the hydrological state could be assumed to be static or assimilated in the linear drift term used in our filter procedure.

The concentration distribution arising after imbibition 1 led to electro-diffusion signals ($SP^{diff}$). When surface charge effects can be neglected (i.e. large pore size), the electro-diffusion source currents can be calculated from the ionic mobilities in the pore water. In this study, this contribution was found to be rather minor. For other types of pore water chemistry, this signal can reach several tenths of mV (Maineult et al., 2004). The second major influence of the water chemistry is differences in chloride concentration in the vicinity of the electrode surfaces. We find for our experimental data that rather small differences in chloride concentration (a 1.26 ratio) yield a higher electrode effect than $SP^{EK}$. The fact that $SP^{EK}$ and $SP^{elec}$ can have approximately the same amplitudes highlights that the chloride concentration effects upon the electrodes cannot be neglected. Indeed, they could either mask (Fig. 10j) or enhance (Fig. 11i and j) the measured SP signal. This implies that the electrode effects have to be explicitly modeled or filtered out when interpreting SP monitoring results, even in apparently well-controlled laboratory conditions.

After correction for temperature effects and chloride concentrations, there are still some residual long-term drifts of the electrodes that cause a certain spread in the observed SP data with time (Fig. 9c). This effect might be partly related to unaccounted concentration differences. However, we cannot exclude that irreversible electrode processes occur at these time-scales that might affect the equilibrium electrode potential. Notably, how does the possible accumulation of Ag or AgCl on electrodes due to electrode reactions affect these potentials? This subject is outside the scope of the present study, but would be worthwhile to revisit in the future.

*5.2 Comparison with previous studies using Ag-AgCl electrodes*

This work highlights that transport simulations are needed to evaluate the chloride concentration effects on the electrode potential. In well-controlled conditions, this results in a fairly accurate reproduction of the measured signals. If the electrodes are in chambers, their modeling is more challenging, as an explicit model of the evolution of the chamber fluid is needed that considers the dimensions of the chambers and the diffusion through the porous interface to the pore water solution. This can be problematic in partially saturated media as the exchange between the chamber and the geological medium can vary from one electrode to another (e.g., by saturation-dependent diffusion coefficients).

Different types of corrections have been used in the past to remove electrode effect. Linde et al. (2007) assigned a general shift of the SP data to assure zero values at the end of drainage. Another common correction approach is to correct for assumed linear drifts. In a sand column drainage experiment, Mboh et al. (2012) processed their SP data to correct for a linear drift between the beginning and the end of the drainage. We considered the raw data by Mboh et al. (2012), their figure 8. We shifted all their data to 0 mV in the beginning of the experiment before drainage started (Fig. 12a). Prior to any water movement and during steady-state saturated flow (i.e., the first hour), no clear long-term electrode drift can be seen in their data. At the end of the drainage (10 h in Fig. 12a), the authors observe that the SP signals do not tend to zero because of contributions other than streaming potentials. These residual SP values are plotted against the matrix potential in Fig. 12b. The residual SP signals are of the same magnitude as those related to the peak of the streaming potential and they seem related to the matric potential in the column and thus to saturation. For this experiment, Mboh et al. (2012) used tap water, which in general contains a low chloride concentration (e.g., in Switzerland, the recommended value for water distribution is $C_{Cl^-} = 5.64 \times 10^{-4}$ mol L$^{-1}$). Considering that the temperature is constant along the column (20.1°C) and using Eq. [10], the residual SP signals after drainage could be explained with a ratio between the activity at vicinity of the $i^{th}$ electrode and the one at the reference electrode as small as: $\{Cl^-\}_i / \{Cl^-\}_{ref} \in [0.995\,;1.094]$. Such a ratio could be caused by differential diffusion processes between the medium and the electrode chambers due to differences in saturation. Even if the evolution of $SP^{elec}$ is varying between these experiments, it appears that neither a constant offset nor a linear trend removal can adequately correct for the $SP^{elec}$ effects (c.f., Figs. 10 and 11).

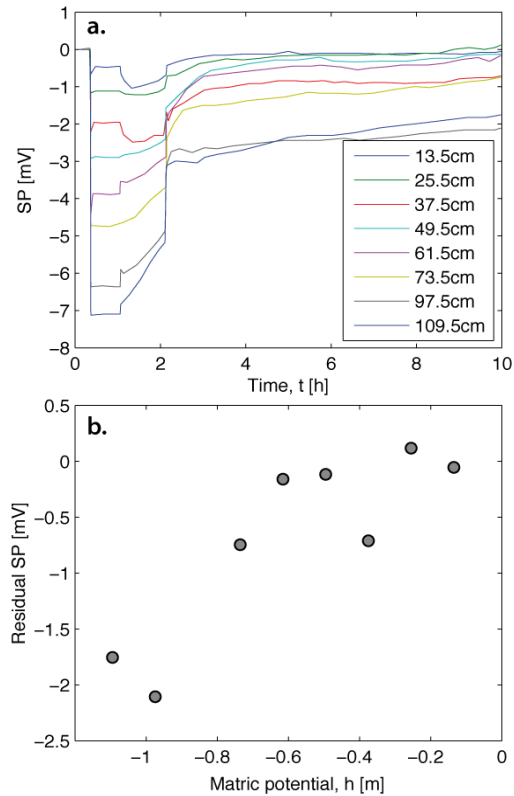

*Fig. 12. (a) Raw SP signal from Mboh et al. (2012) shifted to 0 mV at t = 0 h (modified from their Fig. 8). (b) Residual SP values as a function of matric potential at the end of the drainage experiment t = 10 h.*

Allègre et al. (2010) monitored SP during drainage in a sand-filled column. They used deionized water to increase the values of $SP^{EK}$. After reaching a geochemical equilibrium at ~ 100 µS cm$^{-1}$, they placed the SP electrodes in chambers filled with deionized water, which implies a strong chemical disequilibrium with the electrode. According to Raynauld and Laviolette (1986), the lack of chloride ions in the direct vicinity of the electrodes (i.e., the chamber) induce the dissolution of AgCl(s), which changes the electrode behavior and creates irreversible equilibrium electrode potential differences. The exact behavior of this type of experimental set-up is difficult or even impossible to predict (Janz and Ives, 1968) given the lack of information about the chloride activity within the electrode chamber or the chloride diffusion from the electrode vicinity of the geological medium. Nevertheless, electrode effects are in our minds the most likely reasons for the large residual SP signals that they measured under no-flow conditions, which implies that the explanation of Allègre et al. (2012) in terms of water fluxes is highly unlikely. Electrode effects are also a plausible explanation for why their responses are quite different from those obtained in other experiments (e.g., Linde et al., 2007; Mboh et al., 2012; this work).

## 5.3 What electrodes to use for laboratory purposes?

Given the results of this and previous studies, the question arises about what electrode designs to choose for SP monitoring in partially saturated porous media. Different approaches have been proposed when using Ag-AgCl electrodes to conduct SP measurements for vadose zone studies: electrodes inserted in a chamber filled with deionized water (Guichet et al., 2003; Allègre et al., 2010), with pore water (Mboh et al., 2012) or by directly inserting them in the porous medium (Linde et al., 2007; this work).

To use electrodes inserted in a chamber, one should preferably fill the chamber with the corresponding electrode anion in solution to avoid electrode dissolution and limit electrode effects (see section 5.2 and Mboh et al., 2012). From a theoretical point of view (Eq. [9]) and literature studies (e.g. Junge, 1990; Tallgreen et al., 2005), the most stable electrodes would be those inserted in chambers containing a high concentration of chloride that helps to buffer concentration changes in the electrode vicinity. However, a high concentration in the chamber yields strong diffusion effects (section 3.1) that may result in a pollution of the porous medium under study. To limit these diffusion effects, one could design electrode chambers with reduced surface area (e.g., Snyder et al., 1999; Shao and Feldman, 2007). But at low saturation, this might create contact resistance problems between the electrode and the geological medium. Another solution is to use a very detailed characterization of the electrode chamber and the diffusion process (cf., the test proposed in section 3.1) and to include the electrode leakage within the system that is modeled. It appears that the electrode design used by Maineult et al. (2004, 2005, and 2006) based on Cu-CuSO$_4$ gives very stable results in fully saturated media (10 days in Maineult et al., 2004). The 8 mm thick and 2.1 mm-diameter porous ceramic used as contact with the geological media appears to significantly reduce electrode leakage. To the best of our knowledge, no tests similar to those presented in Fig. 3 have been published for these types of electrodes.

Another option is the one used in this study, in which we directly insert the electrode in the porous medium and simulate the chloride transport, as well as the water flux. The decomposition of the measured signals allowed us to highlight the importance of each contribution to the measured SP (Fig. 10 and 11). This approach demands a rather extensive knowledge of the different processes occurring in the medium and exhaustive characterization of the medium properties, which can be challenging for field applications.

## 6. Conclusions

Electrode effects have been largely ignored in most SP studies concerned with vadose zone applications. This has led to the introduction of theoretically invalid ad hoc theory based on data that have been improperly filtered. From the results presented in this study, we showed that electrode effects can be modeled considering flow, transport and electrical processes in the numerical simulations. We obtain overall satisfying results when comparing the simulations to experimental data from the drainage/imbibition cycles of a sand-filled column.

It appears absolutely vital to consider both flow and transport, together with explicit modeling of electrodes if using chambers, and meticulously controlled conditions if one is to use the resulting SP data to test new theoretical models. Electrodes that are placed directly in the medium are easy to model and results presented herein are promising, while chamber electrodes are interesting if the electrode system is explicitly modeled. It is also clear that the electrolyte in the vicinity of electrodes of the second kind must contain the anion of the used salt in order to function properly, which discards the use of deionized water in the electrode chambers. Electrode effects appear inevitable in longer-term SP monitoring, but this is not necessarily a problem if they can be properly predicted. The results presented herein raises concerns about the ability to monitor evaporative fluxes in a quantitative sense due to associated salinity gradients that are likely to provide electrode-related signals that may be much larger than those due to water fluxes.

**Acknowledgments**

Damien Jougnot thanks Manuel A. Méndez, Philippe Leroy, and Marina Rosas Carbajal for fruitful discussions. Many discussions with André Revil on related subjects have helped to shape this work. Claude Doussan and Dominique Renard provided the sand water retention characterization. Constructive reviews by Chris Hubbard and an anonymous reviewer helped to improve the manuscript.